\documentclass[journal=jctcce,manuscript=letter]{achemso}
\setkeys{acs}{articletitle = true,
chaptertitle = true}

\usepackage[version=3]{mhchem}
\usepackage{atbegshi}
\usepackage{placeins}
\usepackage{indentfirst}
\usepackage{xurl}
\usepackage[colorlinks=true,
allcolors=blue]{hyperref}
\usepackage{graphicx}
\usepackage{svg}
\usepackage{xcolor}
\usepackage[normalem]{ulem}
\newcommand{\add}[1]{{#1}}
\newcommand{\del}[1]{}

\AtBeginDocument{\addtocounter{page}{-1}\AtBeginShipoutNext{\AtBeginShipoutDiscard}}

\graphicspath{{figures/}}
\makeatletter
\providecommand\@mcitecorrectmaxwidthsubitem{0pt}
\makeatother

\author{Mingwei Ding}
\altaffiliation{These authors contributed equally to this work.}
\affiliation{School of Artificial Intelligence and Data Science,
University of Science and Technology of China,
Hefei 230026,
P.R.~China}
\alsoaffiliation{Suzhou Institute for Advanced Research,
University of Science and Technology of China,
Suzhou 215123,
P.R.~China}

\author{Chen Huang}
\altaffiliation{These authors contributed equally to this work.}
\affiliation{School of Artificial Intelligence and Data Science,
University of Science and Technology of China,
Hefei 230026,
P.R.~China}
\alsoaffiliation{Suzhou Institute for Advanced Research,
University of Science and Technology of China,
Suzhou 215123,
P.R.~China}

\author{Yibo Hu}
\affiliation{School of Artificial Intelligence and Data Science,
University of Science and Technology of China,
Hefei 230026,
P.R.~China}
\alsoaffiliation{Suzhou Institute for Advanced Research,
University of Science and Technology of China,
Suzhou 215123,
P.R.~China}

\author{Yifan Li}
\affiliation{Department of Chemistry,
Princeton University,
Princeton,
NJ 08544,
USA}

\author{Zitian Lu}
\affiliation{School of Artificial Intelligence and Data Science,
University of Science and Technology of China,
Hefei 230026,
P.R.~China}
\alsoaffiliation{Suzhou Institute for Advanced Research,
University of Science and Technology of China,
Suzhou 215123,
P.R.~China}

\author{Xingtai Yu}
\affiliation{Division of Life Sciences and Medicine,
University of Science and Technology of China,
Hefei 230026,
P.R.~China}

\author{Duo Zhang}
\affiliation{AI for Science Institute,
Beijing 100080,
P.R.~China}
\alsoaffiliation{DP Technology,
Beijing 100080,
P.R.~China}
\alsoaffiliation{Academy for Advanced Interdisciplinary Studies,
Peking University,
Beijing 100871,
P.R.~China}

\author{Wenxi Zhai}
\affiliation{Shanghai Engineering Research Center of Molecular Therapeutics and New Drug Development,
School of Chemistry and Molecular Engineering,
East China Normal University,
Shanghai 200062,
P.R.~China}
\alsoaffiliation{Shanghai Innovation Institute,
Shanghai 200003,
P.R.~China}

\author{Tong Zhu}
\affiliation{Shanghai Engineering Research Center of Molecular Therapeutics and New Drug Development,
School of Chemistry and Molecular Engineering,
East China Normal University,
Shanghai 200062,
P.R.~China}
\alsoaffiliation{Shanghai Innovation Institute,
Shanghai 200003,
P.R.~China}
\email{tzhu@lps.ecnu.edu.cn}

\author{Qiangqiang Gu}
\affiliation{School of Artificial Intelligence and Data Science,
University of Science and Technology of China,
Hefei 230026,
P.R.~China}
\alsoaffiliation{Suzhou Institute for Advanced Research,
University of Science and Technology of China,
Suzhou 215123,
P.R.~China}
\alsoaffiliation{Suzhou Big Data \& AI Research and Engineering Center,
Suzhou 215123,
P.R.~China}
\email{guqq@ustc.edu.cn}

\author{Jinzhe Zeng}
\affiliation{School of Artificial Intelligence and Data Science,
University of Science and Technology of China,
Hefei 230026,
P.R.~China}
\alsoaffiliation{Suzhou Institute for Advanced Research,
University of Science and Technology of China,
Suzhou 215123,
P.R.~China}
\alsoaffiliation{Suzhou Big Data \& AI Research and Engineering Center,
Suzhou 215123,
P.R.~China}
\email{jinzhe.zeng@ustc.edu.cn}

\title{Automating Computational Chemistry Workflows via OpenClaw and Domain-Specific Skills}

\begin{document}

\begin{tocentry}
\centering
\includegraphics[width=\linewidth]{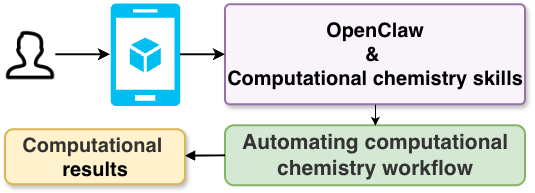}
\end{tocentry}

\begin{abstract}
This work presents a decoupled framework for multi-step computational chemistry automation built on OpenClaw.
\del{OpenClaw, a general-purpose agent, provides general-purpose}\add{OpenClaw serves as the general-purpose agent for} task coordination and supervision\del{;}\add{.}
\del{planning}\add{Planning} skills externalize task descriptions into executable task specifications\del{;}\add{,}
domain skills provide computational chemistry procedures\del{;}\add{,}
and the DPDispatcher skill grounds computation in heterogeneous HPC environments.
In a methane-oxidation reactive MD case study,
the framework coordinated cross-tool execution,
supported bounded recovery from runtime failures,
and extracted reaction networks.
\end{abstract}

\section{Introduction}
Computational chemistry has become an indispensable tool for probing,
predicting,
and rationalizing chemical and materials systems.
In practice,
however,
the major bottleneck is often not the lack of computational methods themselves,
but the difficulty of reliably coordinating heterogeneous software tools,
data representations,
and execution environments for multi-step research tasks.
Real applications routinely require molecular structure preparation,
input generation,
parameter configuration,
data-format conversion,
remote job submission,
monitoring,
error recovery,
and downstream analysis across multiple codes and platforms.
Because these operations are strongly interdependent,
automation depends not only on executing individual tools,
but also on maintaining consistency across task specification,
software invocation,
intermediate data handoff,
and HPC execution.
As such tasks become increasingly multi-step and data-intensive,
the central challenge is no longer simply to automate isolated calculations,
but to achieve robust,
reusable,
and auditable end-to-end automation.\cite{Wu_SmallMethods_2025_v9_pe2500308}

One major line of prior work approaches this problem through workflow-based automation.
Representative examples include general workflow infrastructures such as AiiDA\cite{Pizzi_ComputMaterSci_2016_v111_p218,Uhrin_ComputMaterSci_2021_v187_p110086},
FireWorks\cite{Jain_ConcurrComput_2015_v27_p5037},
and QMflows\cite{Zapata_JChemInfModel_2019_v59_p3191},
as well as domain-oriented workflow systems for specific scientific tasks.
In artificial intelligence (AI)-driven atomistic simulation,
for example,
DP-GEN\cite{Zhang_ComputPhysCommun_2020_v253_p107206} and NepTrain\cite{Chen_ComputPhysCommun_2025_v317_p109859} automate active-learning cycles for machine-learning potentials\del{;}\add{.} CatFlow\cite{Liu_JPhysChemC_2025_v129_p1089} and DP-TI\cite{dpti} support free-energy and thermodynamic-integration workflows\del{;}\add{.} ChecMatE\cite{Guo_JChemPhys_2023_v159_p094801},
AutoSteper\cite{Liu_Carbon_2023_v213_p118180},
APEX\cite{Li_npjComputMater_2025_v11_p88},
and RiD-kit\cite{Fan_BMCMethods_2025_v2_p14} provide workflow support for machine-learning-potential-based materials and molecular simulations\del{; and}\add{. Tools such as} mech2d\cite{Wang_MolBaselSwitz_2023_v28_p4337} and maptool\cite{Wang_ResWashingtonDC_2022_v2022_p9857631} illustrate high-throughput materials-screening workflows.
These systems have proven highly effective for repetitive,
well-defined,
and high-throughput computational tasks by formalizing task dependencies,
provenance tracking,
data organization,
and job execution across heterogeneous computing resources.
However,
workflow-based automation primarily relies on predefined task graphs and execution logic.
Even when limited runtime dynamism is supported,
these approaches become harder to extend when tasks require cross-software adaptation,
context-dependent decision making,
or recovery from unforeseen runtime failures.\cite{Jain_ConcurrComput_2015_v27_p5037,Ganose_DigitDiscov_2025_v4_p1944,Bai_FuturGenerComputSyst_2024_v152_p112}

Recent workflow-centric automation frameworks for scientific model construction and training further underscore the value of modular tool integration and iterative execution in domain-specific settings.\cite{Chen_arXiv_2025_p2506.01868}
In parallel, large language model (LLM)-based agent systems have opened an important complementary direction for scientific automation by coupling natural-language reasoning with external tools, retrieval, and multi-step execution.\cite{Ramos_ChemSci_2025_v16_p2514,Karpas_arXiv_2022_p2205.00445,Yao_arXiv_2023_p2210.03629,Schick_arXiv_2023_p2302.04761}
Rather than relying solely on predefined procedural logic, these systems can interleave reasoning, tool invocation, intermediate-state inspection and evidence gathering, and task-level replanning, thereby adapting to evolving execution contexts during runtime.\cite{Karpas_arXiv_2022_p2205.00445,Yao_arXiv_2023_p2210.03629,Schick_arXiv_2023_p2302.04761}
Recent studies relevant to chemistry and scientific automation, including ChemCrow and CACTUS, suggest that LLM-based agents can coordinate domain tools, scientific knowledge, and execution constraints across complex research workflows.\cite{MBran_NatMachIntell_2024_v6_p525,McNaughton_AcsOmega_2024_v9_p46563,Xiao_arXiv_2026_p2601.01357,bai2026elagentegraficostructured}
This emerging direction is important because it extends automation beyond fixed procedural pipelines toward execution strategies that can respond dynamically to intermediate results, tool outputs, and changing task context.\cite{MBran_NatMachIntell_2024_v6_p525,McNaughton_AcsOmega_2024_v9_p46563,Karpas_arXiv_2022_p2205.00445,Yao_arXiv_2023_p2210.03629}

Current approaches to computational chemistry automation occupy two largely separate design regimes.
Workflow systems provide explicit execution structures and,
in some cases,
robust scheduler integration,
but they typically rely on predefined control flow and limited recovery semantics.\cite{Pizzi_ComputMaterSci_2016_v111_p218,Jain_ConcurrComput_2015_v27_p5037,Zapata_JChemInfModel_2019_v59_p3191,Ganose_DigitDiscov_2025_v4_p1944}
Computational chemistry focused agentic systems provide greater runtime flexibility through dynamic reasoning and tool selection.\cite{MBran_NatMachIntell_2024_v6_p525,McNaughton_AcsOmega_2024_v9_p46563,Zou_Matter_2025_v8_p102263,Lala_arXiv_2023_p2312.07559,Ghafarollahi_AdvMaterDeerfieldBeachFla_2025_v37_pe2413523,Boiko_Nature_2023_v624_p570,choi2026elagenteestructuralartificially,gustin2026elagentecuanticoautomating}
However, in many such systems, planning routines, tool use, and execution interfaces are implemented together within specialized agent stacks, making workflow adaptation and recovery mechanisms less independently replaceable.
As a result,
extending capability may require redesigning agent-side orchestration rather than simply replacing an executable module.
These contrasts can be examined in terms of workflow explicitness,
recovery model,
HPC grounding,
module-replacement granularity,
and dependence on computational chemistry specific design.
From this perspective,
the central limitation is not insufficient reasoning capability,
but engineering entanglement: the tight coupling of reasoning,
workflow specification,
domain execution,
and infrastructure adaptation,
which prevents them from being independently replaced,
extended,
and maintained.

More broadly,
current automation systems can be viewed along a spectrum from scripted workflows,
to workflows with LLM-enabled nodes,
to specialized agents whose task logic is largely embedded in prompts,
planning routines,
and tool-routing policies.\cite{Karpas_arXiv_2022_p2205.00445,Yao_arXiv_2023_p2210.03629,Schick_arXiv_2023_p2302.04761}
Across this spectrum,
reusable tool access does not by itself provide task-specific execution competence:
a general agent may have access to broad capabilities through runtime-loaded tools or skills,
yet still require an explicit task specification for how those capabilities should be combined to achieve a concrete computational objective.\cite{Lala_arXiv_2023_p2312.07559,shao2024assistingwritingwikipedialikearticles,ma2024sciagenttoolaugmentedlanguagemodels}
At the same time,
computational chemistry places unusual pressure on software-facing execution layers,
because executable substrates change frequently across software packages,
file formats,
instrumentation interfaces,
and,
in many practical settings,
HPC environments.\cite{MBran_NatMachIntell_2024_v6_p525,Boiko_Nature_2023_v624_p570,Prince_npjComputMater_2024_v10_p251,chiang2024llamplargelanguagemodel}
These considerations motivate a decoupled design in which general-purpose control is separated from computational chemistry execution.
Rather than embedding domain procedures directly into a specialized agent stack,
we organize automation around a general-purpose agent that handles coordination,
state tracking,
and decision making,
while delegating concrete scientific operations to reusable skills.
In this formulation,
new domain capabilities are introduced by adding or replacing executable skills rather than redesigning the agent itself,
which is especially attractive for computational chemistry.

Here,
we build a computational chemistry automation system on top of the OpenClaw agent-skill framework.
\del{OpenClaw\cite{Steinberger_OpenClaw_2025} provides}\add{OpenClaw\cite{Steinberger_OpenClaw_2025} serves as} the general-purpose control and coordination mechanism\del{;}\add{.} \del{schema-defined}\add{Schema-defined} planning skills convert computational chemistry task descriptions into executable task specifications\del{;}\add{,} domain skills encapsulate computational chemistry procedures and software-facing operations\del{;}\add{,} and DPDispatcher\cite{Yuan_JChemInfModel_2025_v65_p12155} supplies a scheduler-aware execution interface for heterogeneous HPC environments.
In this way,
new computational chemistry capabilities can be introduced,
replaced,
and reused at the skill level without redesigning the agent itself.

In this work,
we demonstrate this design through a case study on methane oxidation reaction exploration,
adapted from our previous molecular dynamics (MD) study of combustion chemistry using a neural network potential.\cite{Zeng_NatCommun_2020_v11_p5713} Starting from a natural language instruction,
the system constructs an explicit task specification and carries out molecular optimization,
file format conversion,
initial reactive system construction,
molecular dynamics simulation,
HPC dispatch and monitoring,
failure recovery,
and trajectory analysis.
This case illustrates that computational chemistry procedures can be externalized into reusable skills,
providing a practical basis for computational chemistry automation.
To support reuse and reproducibility,
we have also released an open-source library for computational chemistry skills,
covering quantum chemistry,
molecular dynamics,
machine-learning potentials,
and molecular representation.

\section{Methodology and Architecture}

\subsection{Design Principles and System Organization}

Our implementation is motivated by a simple principle: computational chemistry automation should externalize not only reusable computational procedures,
but also the task-specific logic that determines how those procedures are organized and executed.
In our formulation,
reusable skills provide domain operations together with usage descriptions available at runtime,
while a task manifest converts a computational chemistry task description into an explicit specification of which skills to combine,
in what order,
and under what conditions.
Under this design,
the agent does not require computational chemistry specific fine-tuning or a deeply specialized internal policy.
Instead,
domain appropriate behavior emerges from the runtime combination of a general agent,
reusable skills,
and task specific guidance.

\begin{figure*}[!t]
\centering
\includegraphics[width=0.94\linewidth]{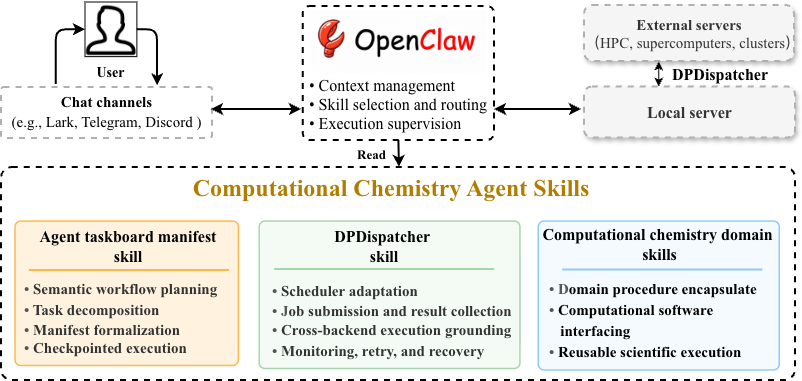}
\caption{System organization of our OpenClaw-based computational chemistry automation system.}
\label{fig:architecture}
\end{figure*}

We instantiate this design using OpenClaw as the general agent framework,
augmented with reusable skills for task coordination and computational chemistry execution,
as shown in Figure~\ref{fig:architecture}.
Rather than constructing a specialized computational chemistry agent from scratch,
we build on a general agent framework maintained by a broader software community,
thereby improving software sustainability while allowing domain experts to focus on task specific capabilities.
At the same time,
a general agent alone is insufficient for computational chemistry automation,
because it does not natively encode scheduler semantics,
scientific data formats,
or software-specific execution procedures.
We therefore place these scientific and execution related capabilities in skills that are loaded at runtime and can be revised independently of the core agent.

To make the decoupled skill architecture practical for computational chemistry automation,
the execution layer must support isolated deployment and reproducible tool invocation.
It must also provide a consistent command-line interface (CLI) so that skills can be executed uniformly at runtime.
This uniform interface simplifies runtime invocation and error handling within the OpenClaw control loop.
Most skills in our ecosystem are therefore built around the uv toolchain.
In particular, uvx invokes tools in temporary isolated environments without requiring global installation,
which reduces dependency conflicts and minimizes interference with the host environment.
This allows each skill to be represented as an executable command together with its required packages,
so that dependencies can be resolved at runtime while remaining modular at the skill level.
Because uv also caches previously built dependencies,
subsequent executions can reuse local packages,
improving deployment efficiency and supporting reproducible workflow execution.
Skills can further constrain tool resolution to specific packages or explicit versions,
thereby improving consistency across runs.
Taken together,
these features provide isolated,
reproducible,
and modular execution environments for skills without entangling dependency management with the core agent.

The role of skills is not to make decisions or act autonomously,
but to serve as structured capability packages containing instructions,
templates,
and validation logic.
In OpenClaw,
relevant skills are loaded into the model context as needed during task execution rather than being exhaustively exposed in advance.
The language model then reasons over the user goal,
the current state,
the loaded skills,
and the execution feedback to determine the next action,
which OpenClaw then executes.
Tool outputs,
generated files,
and external program feedback are returned to the same control loop to inform the next decision step.
Consequently,
high-level planning,
HPC submission,
and computational chemistry operations are all orchestrated by OpenClaw rather than performed independently by isolated modules.
The skills,
in turn,
contribute capability descriptions and procedures that this loop can invoke,
inspect,
and revise.

From this perspective,
our implementation is best understood as an OpenClaw-based general agent system extended with an extensible computational chemistry skill ecosystem.
Some skills focus on workflow coordination,
some on execution grounding,
and some on domain operations,
but they all enter the system through the same mechanism of context loading,
model reasoning,
tool execution,
and feedback.
This design combines the maintainability of a general agent framework with the task specificity required for computational chemistry.

\FloatBarrier

\subsection{Runtime Control Loop and Skill Invocation}

\begin{figure*}[!t]
\centering
\includegraphics[width=0.94\linewidth]{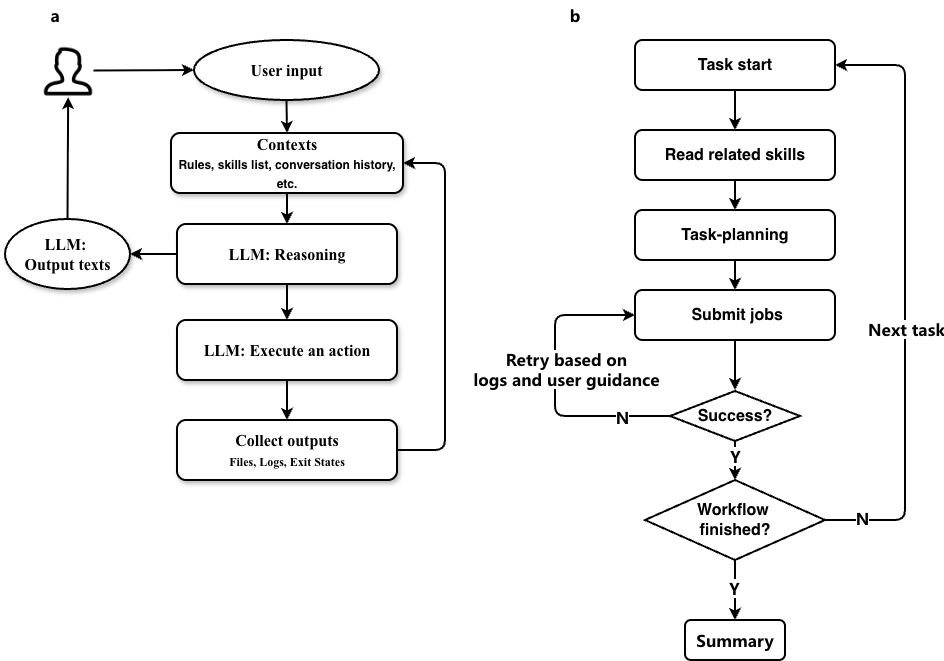}
\caption{(a) Schematic of the LLM-based decision-making process; (b) Workflow of the automated computational task execution process.}
\label{fig:execution_flow}
\end{figure*}

At runtime,
OpenClaw maintains session context that includes the user request,
tool access,
and state tracking across long tasks.
For each step,
it first reads the current conversation state together with the outputs of previous actions.
The language model then produces the next action,
which may involve loading a skill that matches the current need,
asking the user a question,
calling a tool,
generating a workflow artifact,
or executing a domain-specific script.
OpenClaw validates and executes that action,
captures standard output,
error output,
generated files,
and tool results,
and appends them to the active context for the next turn.
This loop continues until the task is completed or human intervention is required.
This process is illustrated schematically in Figure~\ref{fig:execution_flow}a.

The detailed logical flow of the automated execution engine is shown in Figure~\ref{fig:execution_flow}b. Upon task initiation, the system identifies and retrieves the necessary skill sets. This is followed by a task-planning phase in which the LLM decomposes the high-level request into a series of executable jobs. During the execution stage, the system handles job submission to the remote environment and monitors its status. A key feature of this workflow is the autonomous error-recovery loop: when execution does not reach a successful state, the agent retries by analyzing execution logs and, if necessary, requesting user guidance to resolve the bottleneck. This cycle repeats for each sequential task within the workflow manifest until all dependencies are satisfied, ultimately generating a final summary of the research outputs.

\FloatBarrier

Progressive skill loading is particularly advantageous in computational chemistry;
it keeps the context window focused on the immediate task,
thereby avoiding the overhead of processing documentation for every supported software package simultaneously.
This modularity also facilitates system expansion,
as integrating new software typically requires only adding or updating specific skills rather than modifying the core agent logic.
Such decoupling motivates the adoption of a general-purpose agent framework over an architecture tailored exclusively to computational chemistry.
Under this design,
the core infrastructure leverages broader software maintenance and community development,
while scientific expertise remains concentrated within reusable,
domain-specific skills.

\subsection{The Agent Taskboard Manifest and the DPDispatcher Skill}

Two skills are especially important in our implementation.
The first converts an open-ended computational task description into an explicit, executable workflow representation,
thereby bridging the gap between broad capability access and task-specific execution competence,
whereas the second grounds long-running tasks in real execution environments.

The Agent Taskboard Manifest skill addresses a key limitation of purely tool-augmented general agents.
While dynamic skill loading endows the model with a wide range of executable capabilities,
tool access alone does not specify how those capabilities should be ordered and combined to fulfill a concrete computational task description.
In traditional dedicated agents, such task-specific execution competence is typically embedded directly in the agent through carefully engineered prompt constraints that channel the model along a fixed procedural path.
To retain generality without compromising execution rigor,
we instead implement workflow coordination as a specialized skill.
This skill serves as a meta-prompting layer that generates task-guidance prompts and execution constraints dynamically at runtime.

Specifically, it translates the user-provided computational task description into a structured task specification that explicitly defines stages, dependencies, required inputs, expected outputs, and validation conditions. To control reasoning complexity and manage the language model context window, the resulting workflow description adopts a lazy-loading strategy in which only the currently active subtask is introduced into the agent context. This explicit and modular representation improves inspectability, facilitates state-dependent recovery, and provides a general agent with the targeted procedural constraints that are otherwise hardcoded in dedicated expert systems. Because the coordination mechanism is dynamically loaded at runtime in the same manner as other skills, it remains replaceable and can be refined without modifying the underlying OpenClaw runtime.

The DPDispatcher skill implements execution grounding.
Built on DPDispatcher\cite{Yuan_JChemInfModel_2025_v65_p12155}, an open-source package developed by the authors, it supports job-script generation, job submission, status monitoring, and result retrieval across local and remote environments.
In our framework, the corresponding skill translates agent intent into validated DPDispatcher descriptors, including machine settings, resource requests, file-transfer specifications, and task commands.
This design shields the core agent from scheduler-specific syntax while still exposing the operational choices that matter for scientific execution.
Through DPDispatcher, the same agent can target local shell execution, Slurm \cite{Yoo_LectNotesComputSci_2003_p44}, the Portable Batch System (PBS) \cite{Jones_BeowulfClustComputLinux_2001_p369,Nitzberg_BookCharp_2004_p183},
and Load Sharing Facility (LSF) \cite{Zhou_SoftwPrExp_1993_v23_p1305} through a unified interface. Queueing, waiting, monitoring, and result collection are therefore treated as normal workflow states rather than as ad hoc shell-level bookkeeping.
This point is important in computational science, where the practical bottleneck is often not a single command, but the full life cycle of remote execution.

\subsection{Computational Chemistry Agent Skills}

Our domain capabilities are further organized as computational chemistry Agent Skills.
We have released an open-source library for computational chemistry Agent Skills under the LGPL-3.0 license at \url{https://github.com/jinzhezenggroup/computational-chemistry-agent-skills}.
The repository provides reusable skills for common computational chemistry operations and serves as a public starting point for community extension.
\del{A further point is that skills are not in one-to-one correspondence with software packages.
A single skill may coordinate several programs for one task,
and one software package may be exposed through multiple skills that target different operations or levels of abstraction.
For this reason,
Table~\ref{tab:skills_ecosystem} summarizes related software in the current repository rather than implying a one-skill-per-software mapping.}
\add{The skills in this repository are primarily written around computational chemistry software interfaces and common software uses.
Some skills target a specific program or command-line workflow, whereas others route or coordinate related subskills for a broader package family.
Table~\ref{tab:skills_ecosystem} summarizes the software packages and workflow categories currently covered by this skill set.}
At the time of writing,
the repository spans agent workflow,
analysis,
atomistic workflows,
data processing,
machine-learning potentials,
molecular dynamics,
molecular representation,
quantum chemistry,
and tools.
The current skill set spans a broad range of computational chemistry operations,
but remains incomplete and should therefore be viewed as an extensible ecosystem rather than a comprehensive collection.

\add{Because these skills encode software usage rather than chemistry-system definitions, they are intended to be combined with task manifests across different molecular or materials systems.
For example, a LAMMPS-related skill provides the instructions, templates, and checks needed to use LAMMPS in an MD workflow independently of the chemical system, but it may cover only selected LAMMPS operations.
Transferability is therefore strongest when a new task reuses covered software operations, and weaker when it requires an uncovered package, option, or validation criterion.}

\add{Extending the repository follows the same software-interface logic.
A new or expanded skill is developed by cataloging the operations, commands, templates, and validation checks required by the target software, writing a \texttt{SKILL.md} entry point following the agentskills.io specification, adding reference documentation or assets as needed, and testing the skill with OpenClaw.
The repository includes \texttt{create-skill}, an Agent Skill for helping contributors start developing new computational-chemistry software skills.
The accompanying \texttt{CONTRIBUTING.md} guide describes how community contributions can add or extend such skills.}

\begin{table*}[!t]
\centering
\caption{Current category-level view of the skill repository.
The current list is not exhaustive,
and each category may contain multiple skills and software interfaces.}
\label{tab:skills_ecosystem}
\renewcommand{\arraystretch}{1.3}
\begin{tabular}{p{5.0cm} p{10.0cm}}
\hline
\textbf{Category} & \textbf{Related Software} \\
\hline
Analysis & Phonopy\cite{Togo_JPhysCondensMatterInstPhysJ_2023_v35_p353001},
ReacNetGenerator\cite{Zeng_PhysChemChemPhys_2020_v22_p683} \\
Atomistic workflows & ASE\cite{HjorthLarsen_JPhysCondensMatter_2017_v29_p273002} \\
Data processing & Open Babel\cite{OBoyle_JCheminform_2011_v3_p33},
dpdata\cite{Zeng_JChemInfModel_2025_v65_p11497},
Packmol\cite{Martinez_JComputChem_2009_v30_p2157},
pymatgen\cite{Ong_ComputMaterSci_2013_v68_p314} \\
Machine-learning potentials & DeePMD-kit\cite{Zeng_JChemPhys_2023_v159_p054801,Zeng_JChemTheoryComput_2025_v21_p4375},
DP-GEN\cite{Zhang_ComputPhysCommun_2020_v253_p107206} \\
Molecular dynamics & Amber\cite{Case_JChemInfModel_2023_v63_p6183,Case_JChemInfModel_2025_v65_p7835},
LAMMPS\cite{Thompson_ComputPhysCommun_2022_v271_p108171} \\
Molecular representation & RDKit\cite{RDKit},
Uni-Mol\cite{Zhou_ICLR_2023_UniMol} \\
Quantum chemistry & ABINIT\cite{Verstraete_JChemPhys_2025_v163_p164126},
CP2K\cite{Kuhne_JChemPhys_2020_v152_p194103},
DFTB+\cite{Hourahine_JChemPhys_2020_v152_p124101},
Gaussian\cite{g16},
GPAW\cite{Mortensen_JChemPhys_2024_v160_p092503},
Quantum ESPRESSO\cite{Giannozzi_JPhysCondensMatterInstPhysJ_2009_v21_p395502,Giannozzi_JPhysCondensMatterInstPhysJ_2017_v29_p465901},
SIESTA\cite{Garcia_JChemPhys_2020_v152_p204108},
VASP\cite{Kresse_PhysRevB_1993_v47_p558},
\add{xtb\cite{Bannwarth_WiresComputMolSci_2021_v11_pe1493}} \\
Tools & DPDispatcher\cite{Yuan_JChemInfModel_2025_v65_p12155} \\
\hline
\end{tabular}
\end{table*}

\FloatBarrier

\section{Case Study: Methane Oxidation Reactive MD Automation}

\begin{figure*}[!t]
\centering
\includegraphics[width=0.94\linewidth]{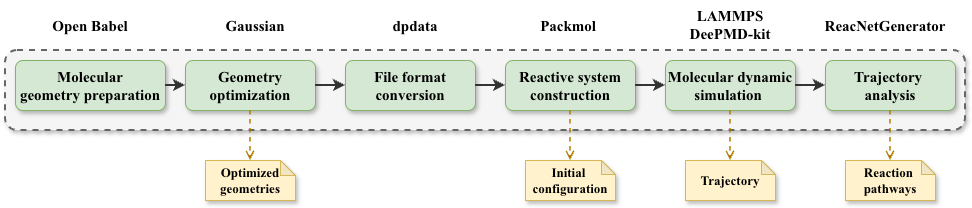}
\caption{Workflow of the methane oxidation MD simulation.}
\label{fig:reactive_md_case}
\end{figure*}

To evaluate the practical feasibility of the proposed architecture for computational chemistry workflows,
we apply it to a case study involving reactive molecular dynamics simulation of methane oxidation and subsequent extraction of reaction pathways from the resulting trajectories, as shown in Figure~\ref{fig:reactive_md_case}.
This task is adapted from our previous MD study of combustion chemistry using a neural network potential,\cite{Zeng_NatCommun_2020_v11_p5713} and is used here to examine whether a general agent equipped with reusable planning and domain skills can orchestrate the corresponding workflow in an explicit, traceable, and recoverable manner.
The case is suitable for this purpose because it requires coordination across multiple interdependent stages,
including molecular structure preparation,
quantum chemical optimization,
mixed-system construction,
reactive molecular dynamics simulation,
and trajectory analysis,
while also depending on interoperability across heterogeneous software packages and execution environments.

In this case,
we submit a natural-language request to OpenClaw,
with \del{GPT-5.2 (xhigh) or GPT-5.4 (xhigh)}\add{GPT-5.5 (xhigh)} serving as the reasoning engine.
\add{The base LLM can have a substantial effect on long-horizon agent behavior.
Earlier model configurations were sufficient for many local tool-use steps but were less reliable at maintaining the full workflow state,
recognizing subtle execution failures,
and applying corrective actions without introducing new inconsistencies.
GPT-5.5 was selected as one of the strongest available agentic reasoning models for the documented run.
This choice does not eliminate errors;
rather,
it improves the probability that the agent follows the loaded skills,
uses execution feedback appropriately,
and recovers from mistakes through tool-grounded correction.}
The prompts are provided in the Supporting Information.
The request defines a multistep methane oxidation workflow that includes generating isolated-molecule structures with Open Babel\cite{OBoyle_JCheminform_2011_v3_p33}, optimizing \ce{CH4} and \ce{O2} at the B3LYP\cite{Stephens_JPhysChem_1994_v98_p11623}/6--31G(d,p)\cite{Hehre_JChemPhys_1972_v56_p2257,Hariharan_TheorChimActa_1973_v28_p213} level of theory, converting the optimized structures into standardized molecular files with dpdata\cite{Zeng_JChemInfModel_2025_v65_p11497}, packing a bulk system containing 50 \ce{CH4} and 100 \ce{O2} molecules at a target density of 0.25~g/cm$^3$ using Packmol\cite{Martinez_JComputChem_2009_v30_p2157}, and generating a LAMMPS-readable data file.
For the reactive molecular dynamics stage,
the agent receives as external input a pretrained Deep Potential\cite{Zhang_BookChap_NIPS_2018_v31_p4436} model for the methane oxidation system, here taken from Ref.~\citenum{Zeng_QuantumChemAgeMachLearn_2023_p279},
and incorporates it into LAMMPS\cite{Thompson_ComputPhysCommun_2022_v271_p108171} through DeePMD-kit\cite{Zeng_JChemPhys_2023_v159_p054801,Zeng_JChemTheoryComput_2025_v21_p4375}.
It then performs short structural relaxation or low-temperature pre-equilibration to remove unfavorable local contacts,
heats the system to 3000~K under the NVT ensemble with the Nose--Hoover thermostat,
and runs a 1~ns reactive molecular dynamics simulation with a time step of 0.1~fs while recording key simulation parameters,
thermodynamic outputs,
and atomic trajectories.
Finally,
ReacNetGenerator\cite{Zeng_PhysChemChemPhys_2020_v22_p683} analyzes the trajectory to track \ce{CH4},
\ce{O2},
radical intermediates,
and oxidation products,
and to extract the dominant methane oxidation pathways and key intermediates.

Rather than mapping this computational task description directly to a fixed procedure,
the framework invokes a schema-defined task-planning skill to externalize the objective into an explicit workflow manifest.
Rather than serving as a simple step list,
this manifest encodes stage dependencies,
completion criteria,
validation conditions,
and execution constraints in a form that can be interpreted at runtime.
For the present case,
the resulting computational workflow comprises six principal stages:
molecular preparation,
geometry optimization,
file format conversion,
reactive system construction,
molecular dynamics simulation,
and trajectory analysis.
In addition, pre-flight validation, scheduler submission, and runtime monitoring are included as workflow-level control steps.
Throughout this process,
OpenClaw maintains the global task state and orchestrates the execution flow,
while reusable domain skills provide the computational chemistry procedures required at each stage.
\add{This case study did not use a methane-oxidation-specific skill; instead, it combined reusable software-interface skills for molecular preparation, file conversion, LAMMPS/DeePMD-kit molecular dynamics, scheduler dispatch, and trajectory analysis.}
As a result,
execution is governed by validated intermediate outputs rather than by a fixed linear script.

\FloatBarrier

Before intensive execution,
the framework performs pre-flight checks on executables,
dependencies,
scheduler access,
and required file paths,
thereby reducing failures arising from environmental mismatches.
Reactive MD execution is then grounded in the target HPC environment through the DPDispatcher skill,
which translates the planned task into scheduler-compatible job descriptors.
In this way,
the orchestration environment and the scientific execution environment remain explicitly connected while operationally separated.

During execution,
task advancement is state-driven rather than purely sequential.
Stage transitions occur only after the required intermediate outputs have been produced and validated.
If an individual stage fails,
the framework inspects the current execution state and applies stage-appropriate bounded recovery actions---such as parameter repair,
retry,
or rollback---before proceeding.
When recovery conditions are not satisfied,
execution halts,
thereby preventing unbounded deviation from the intended task procedure.

We successfully executed the entire workflow in repeated end-to-end demonstrations.
\add{The input files and parameters submitted to each software package were verified to be consistent with the published protocol~\cite{Zeng_NatCommun_2020_v11_p5713}. The generated MD outputs were checked against the established methane-oxidation workflow and showed the expected qualitative consistency with that protocol, including trajectory-derived \ce{CH4} and \ce{O2} consumption and \ce{H2O}, CO, and \ce{CO2} formation as summarized in Supporting Information Figure~S1 and in trace rows 105, 116, and 119 of Table~S1. Normal run-to-run numerical variability in MD trajectories may arise from the underlying simulation engines and execution environment, rather than from the OpenClaw-generated workflow itself.}
Owing to the inherent stochasticity of LLMs,
individual runs exhibited variations in the dialogue trajectories,
the generated action plans,
and the specific errors encountered.
In these demonstrations,
the system leveraged error feedback to resolve such issues
and complete the assigned tasks.
\add{Specifically, this feedback included logs, return codes, schema checks, and generated artifacts.
The framework used these signals together with the explicit workflow state to select bounded recovery actions,
such as retry,
parameter repair,
corrective routing,
or termination.
When the available evidence did not support a defined recovery action,
the workflow halted or requested human guidance.
For the methane-oxidation run reported in detail in Supporting Information Table S1,
the cleaned task-relevant trace consumed 462,662 input tokens,
76,129 output tokens,
and 6,030,336 cache tokens after excluding repeated heartbeat checks and no-op responses. At current GPT-5.5 API pricing, this usage corresponds to an estimated marginal LLM inference cost of approximately US\$7.6, suggesting that language-model orchestration can be modest relative to expert manual setup and debugging once the relevant skills and computational infrastructure are available, while HPC costs and scientific validation remain separate task-dependent costs.
The trace also records a representative recovery event during scheduler-submitted DeePMD/LAMMPS execution,
where an environment-activation incompatibility prevented the initial MD job from completing.
Using scheduler and DPDispatcher logs together with the workflow state,
OpenClaw identified the failure mode,
revised the execution wrapper,
reset the dispatcher state,
and resumed the molecular-dynamics and trajectory-analysis stages.
}

\section{Conclusions}

This work demonstrates multistep computational chemistry automation through a decoupled design built on OpenClaw.
OpenClaw provides general-purpose task coordination and supervision,
planning skills externalize computational chemistry task descriptions into executable task specifications,
the DPDispatcher skill grounds computation in real HPC environments,
and domain skills provide software-specific procedures and result-handling logic within the agent control loop.
Through the methane-oxidation reactive MD case study,
we show that this framework can organize multistage computational chemistry tasks,
coordinate cross-tool execution,
and support bounded recovery from runtime failures without relying on a fully pre-encoded workflow.
These results suggest that general-purpose agents,
coupled with modular skill ecosystems,
offer a scalable and maintainable path toward practical automation across computational chemistry and materials science.
Future work will focus on expanding the reusable skill ecosystem to support a broader range of computational tasks---such as high-throughput materials screening,
periodic solid-state computations,
and active-learning potential generation---while fostering community contributions to maintain the library.
Furthermore,
integrating the agent with rigorous provenance-tracking databases and enabling closed-loop autonomous workflows remain important directions for scientific discovery.\cite{Bai_DigitDiscov_2025_v4_p2123,Bai_NatCommun_2024_v15_p462}

\section*{Data Availability}

\add{All agent skills can be found at \url{https://github.com/jinzhezenggroup/computational-chemistry-agent-skills}.}

\begin{acknowledgement}

This work is supported by
Advanced Materials - National Science and Technology Major Project (No.
2025ZD0618700 to J.Z.)
and USTC Bihe Youth Program for Interdisciplinary Innovation (BH-202520 to C.H.).
Q. G. acknowledges the support from the National Natural Science Foundation of China (Grant No. 12504285) and Natural Science Foundation of Jiangsu Province (Grant No. BK20250472).
\add{The AI-driven simulations were performed on the robotic AI-Scientist platform of Chinese Academy of Sciences.}
Computing resources are provided by
the Supercomputing Center of University of Science and Technology of China
and the Open Source Supercomputing Center of S-A-I.
J.Z.
thanks members of the linux.do community forum for helpful discussions on large language models and OpenClaw.
\add{OpenClaw, with GPT-5.4 and GPT-5.5 as reasoning engines, was used to edit the English in this manuscript.}

\end{acknowledgement}

\begin{suppinfo}

\add{Prompts for the Methane Oxidation Workflow; Methane Oxidation Species Evolution; OpenClaw Interaction Trace and Token Usage for the Methane Oxidation Workflow.}

\end{suppinfo}

\bibliography{combined,tmplib}

\end{document}


\section{Prompts for the Methane Oxidation Workflow}

\begin{verbatim}
For this task your workspace is <path>, we are doing a methane and oxygen
combustion simulation to explore the simulation trajectory. Start by
searching for the SMILES and generating the initial 3D coordinates for both
molecules. After that, optimize the geometry using B3LYP/6-31G(d,p) freq in
Gaussian. This needs to be executed via dpdispatcher on m1 using slurm
queue main, max 1 node, 16 cores, and 8 GB memory, making sure to run source
/group/software/g16.sh first. After optimized and extract coordinates by
dpdata, build the reaction system as a cubic box with 0.25 g/cm3 density
containing 50 units of methane and 100 units of oxygen, with no diluent.
Finally, run the MD simulation using DeePMD via dpdispatcher on m1 (Slurm,
main queue, 1 GPU). Download the potential from
https://github.com/tongzhugroup/Chapter13-tutorial/raw/master/graph.pb
with element order C-H-O. Set units to real and write dump output every
1000 steps. The simulation runs an NVT reactive MD phase at 3000 K for 1 ns
with a 0.1 fs timestep. Final analysis of reaction pathway.

machine_m1.json containing remote_root and remote_profile as part of
dpdispatcher required can help logging in to m1 are placed in your workspace.

Please refine this simple plan into a complete workflow named
ch4_o2_combustion_workflow output in your workspace.
\end{verbatim}

The following \texttt{AGENTS.md} file was available in the workspace.

\begin{verbatim}
# AGENTS.md - Your Workspace

You are an assistant to run computational-chemistry tasks.

Speak English.

Do not do anything you are not asked to.

Plan the workflow (read the plan skill) before conducting a complex workflow.
List the software, skills, and commands you want to use.

Prefer using skills if possible.

Always prefer tmux to run commands, especially long tasks.

Always use dpdispatcher to run each task on remote servers. Do not use `ssh`
manually. Do not visit the server if you are not granted.

Exec note:
- In this environment, do not explicitly pass `host` to the `exec` tool unless
  the user specifically asks.
- Explicit `host=sandbox` may fail with sandbox unavailable.
- Explicit `host=gateway` may fail due to host policy.
- Omitting `host` can still work and should be preferred for normal command
  execution.

Machine information:
- Server n1
  - port 22
  - SSH keys can be automatically found
  - Username: jzzeng
  - work path: /home/jzzeng/dpgen_workdir (this should be strictly followed)
  - Slurm management system, queue: main
  - Software:
    - DeePMD-kit: source /group/software/deepmd-kit-3.1.1/bin/activate
      /group/software/deepmd-kit-3.1.1
    - Gaussian: source /group/software/g16.sh
    - uvx: /home/jzzeng/miniconda3/bin/uvx
- local machine
  - software:
    - uvx: /home/ubuntu/miniforge3/bin/uvx
\end{verbatim}

\section{Methane Oxidation Species Evolution}
\label{sec:species_evolution}

\begin{figure}[!ht]
\centering
\includegraphics[width=0.9\linewidth]{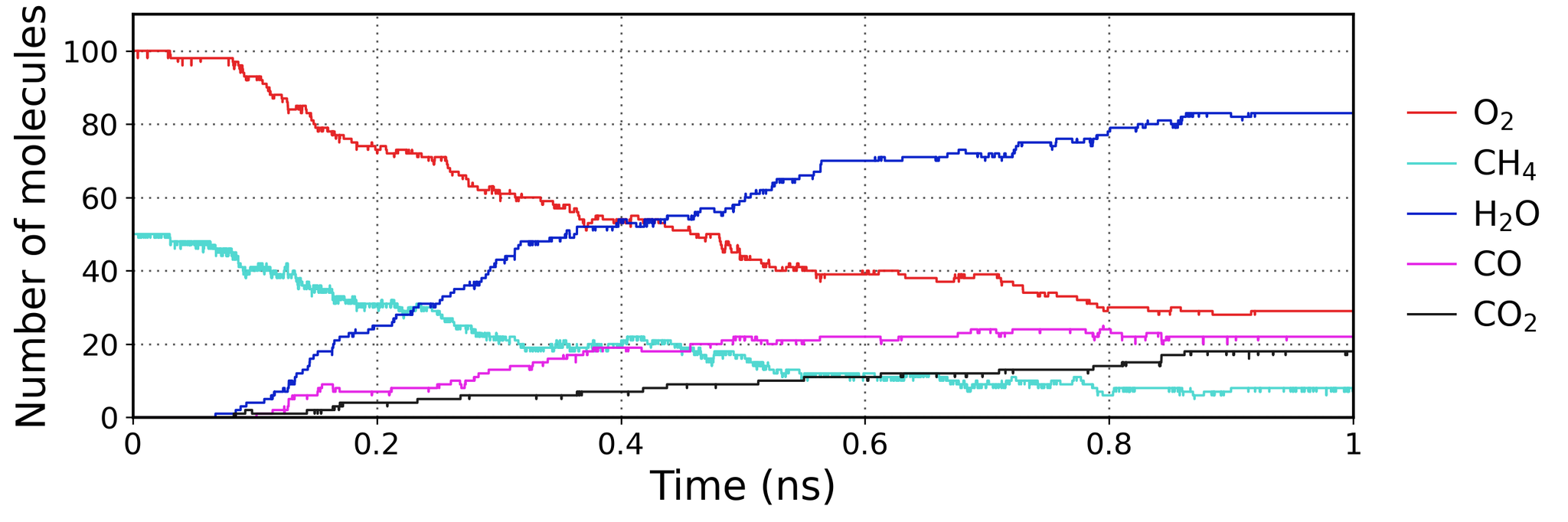}
\caption{Time evolution of the numbers of major molecular species in the methane-oxidation reactive MD trajectory. The trajectory starts from 50 \ce{CH4} and 100 \ce{O2} molecules. During the 1~ns simulation, \ce{CH4} and \ce{O2} are consumed while \ce{H2O}, CO, and \ce{CO2} are formed, providing a qualitative check that the OpenClaw-generated workflow produces the expected methane-oxidation chemistry.}
\label{fig:species_evolution}
\end{figure}

\FloatBarrier

\section{OpenClaw Interaction Trace and Token Usage for the Methane Oxidation Workflow}
\label{sec:interaction_trace}

The interaction trace summarized below was generated with OpenClaw version 2026.4.29 (a448042) using model GPT-5.5 with xhigh reasoning. The table excludes repeated heartbeat or cron-monitoring turns, no-op \texttt{NO\_REPLY}/\texttt{HEARTBEAT\_OK} responses, hidden reasoning blocks, and raw tool-result payloads. Tool-call rows keep the called tool and the executed command, path, URL, query, action, or step list, but omit AI-written intermediate summaries. Passive monitor/log-check turns are omitted unless they led to a diagnosis, repair, report generation, or final scientific response. Token counts are therefore reported for the displayed, task-relevant rows only and are summed directly from the reported usage fields for those rows. The total displayed usage was 462,662 input tokens, 76,129 output tokens, and 6,030,336 cache tokens.

\begin{landscape}
\begingroup
\tiny
\sloppy
\emergencystretch=2em
\setlength{\tabcolsep}{1.5pt}
\setlength{\LTleft}{0pt}
\setlength{\LTright}{0pt}
\setlength{\LTcapwidth}{\linewidth}
\renewcommand{\arraystretch}{1.08}
\begin{longtable}{@{}r >{\RaggedRight\arraybackslash\hspace{0pt}}p{2.15cm} >{\RaggedRight\arraybackslash\hspace{0pt}}p{15.2cm} r r r@{}}
\caption{Cleaned OpenClaw interaction trace for the methane oxidation case study. Tool calls are retained in the format \texttt{tool; command/path/url}.}\label{tab:openclaw_interaction_trace}\\
\hline
\# & Type & Content & Input & Output & Cache \\
\hline
\endfirsthead
\caption[]{Cleaned OpenClaw interaction trace for the methane oxidation case study (continued).}\\
\hline
\# & Type & Content & Input & Output & Cache \\
\hline
\endhead
\hline
\endfoot
1 & \textbf{User message} & For this task your workspace is \texttt{\detokenize{/home/ubuntu/.openclaw/workspace/methane}}, we are doing a methane and oxygen combustion simulation to explore the simulation trajectory. Start by searching for the SMILES and generating the initial 3D coordinates for both molecules. After that, optimize the geometry using B3LYP/6-31G(d,p) freq in Gaussian via DPDispatcher on m1 using Slurm queue main, max 1 node, 16 cores, and 8 GB memory, making sure to run \texttt{\detokenize{source /group/software/g16.sh}} first. After optimization and coordinate extraction by dpdata, build the reaction system as a cubic box with 0.25 g/cm$^3$ density containing 50 units of methane and 100 units of oxygen, with no diluent. Finally, run the MD simulation using DeePMD via DPDispatcher on m1 (Slurm, main queue, 1 GPU). Download the potential from \url{https://github.com/tongzhugroup/Chapter13-tutorial/raw/master/graph.pb}. &  &  &  \\
2 & Agent tool call & read; path: \texttt{\detokenize{openclaw/skills/taskflow/SKILL.md}} & 16,588 & 565 & 0 \\
3 & \textbf{Agent message + tool call} & I'll turn this into a concrete, runnable workflow directory under the methane workspace, using the m1 dpdispatcher config you mentioned and keeping the computational steps separate and reproducible.\newline update\_plan; steps: \texttt{\detokenize{Inspect methane workspace and m1 config; Resolve molecule identifiers and workflow structure; Write workflow files and scripts; Run lightweight validation}} & 1,409 & 628 & 16,384 \\
4 & Agent tool call & exec; command: \texttt{\detokenize{find /home/ubuntu/.openclaw/workspace/methane -maxdepth 2 -type f -o -type d | sort | sed 's#^#/##' | head -200}}\newline exec; command: \texttt{\detokenize{find /home/ubuntu/.openclaw/workspace/methane -maxdepth 2 -name 'machine_m1.json' -type f -print -exec sed -n '1,220p' () \;}} & 511 & 168 & 17,408 \\
5 & Agent tool call & web\_search; query: \texttt{\detokenize{methane canonical SMILES PubChem}}\newline web\_search; query: \texttt{\detokenize{oxygen molecule canonical SMILES PubChem}}\newline exec; command: \texttt{\detokenize{python - <<'PY' import importlib.util mods=['rdkit','ase','dpdata','dpdispatcher'] for m in mods: print(m, bool(importlib.util.find_spec(m))) PY}} & 2,371 & 689 & 15,872 \\
6 & Agent tool call & web\_fetch; url: \traceurl{https://pubchem.ncbi.nlm.nih.gov/rest/pug/compound/name/methane/property/CanonicalSMILES,IsomericSMILES/TXT} & 1,154 & 578 & 17,408 \\
7 & Agent tool call & web\_fetch; url: \traceurl{https://pubchem.ncbi.nlm.nih.gov/rest/pug/compound/name/methane/property/CanonicalSMILES/TXT}\newline web\_fetch; url: \traceurl{https://pubchem.ncbi.nlm.nih.gov/rest/pug/compound/name/oxygen/property/CanonicalSMILES/TXT} & 444 & 119 & 18,432 \\
8 & Agent tool call & read; path: \texttt{\detokenize{skills/run-gauss/SKILL.md}} & 1,865 & 562 & 17,920 \\
9 & Agent tool call & read; path: \texttt{\detokenize{skills/dpdisp-submit/SKILL.md}} & 1,835 & 45 & 18,432 \\
10 & Agent tool call & read; path: \texttt{\detokenize{skills/openbabel/SKILL.md}} & 3,483 & 561 & 19,456 \\
11 & Agent tool call & exec; command: \texttt{\detokenize{uvx --with dpdispatcher dargs doc dpdispatcher.entrypoints.submit.submission_args | sed -n '1,240p'}} & 4,029 & 1,314 & 19,968 \\
12 & Agent tool call & exec; command: \texttt{\detokenize{python - <<'PY' import dpdispatcher, inspect, pkgutil, os, json print('dpdispatcher', dpdispatcher.__version__ if hasattr(dpdispatcher,'__version__') else dpdispatcher) print(os.path.dirname(dpdispatcher.__file__)) PY}} & 1,079 & 1,129 & 23,552 \\
13 & Agent tool call & exec; command: \texttt{\detokenize{python - <<'PY' from dpdispatcher import Machine, Resources, Task, Submission import inspect for cls in [Machine, Resources, Task, Submission]: print('\nCLASS', cls) print(inspect.signature(cls.__init__)) PY}} & 697 & 94 & 24,064 \\
14 & Agent tool call & exec; command: \texttt{\detokenize{python - <<'PY' import dpdispatcher.entrypoints.submit as s import inspect print(inspect.getsource(s)) PY}} & 1,507 & 66 & 23,552 \\
15 & Agent tool call & exec; command: \texttt{\detokenize{python - <<'PY' import dpdispatcher, pkgutil for m in pkgutil.walk_packages(dpdispatcher.__path__, dpdispatcher.__name__+'.'): if 'entry' in m.name or 'submit' in m.name or 'cli' in m.name: print(m.name) PY}} & 602 & 103 & 24,576 \\
16 & Agent tool call & exec; command: \texttt{\detokenize{python - <<'PY' from dpdispatcher.entrypoints import submission import inspect print(inspect.getsource(submission)[:4000]) PY}} & 760 & 70 & 24,576 \\
17 & Agent tool call & exec; command: \texttt{\detokenize{python - <<'PY' import os, dpdispatcher root=os.path.dirname(dpdispatcher.__file__) for dirpath, dirs, files in os.walk(root): for f in files: if f.endswith('.py'): p=os.path.join(dirpath,f) txt=open(p,encoding='utf-8').read(errors='ignore') if 'submission_args' in txt or 'dargs' in txt: print(p) PY}} & 1,018 & 142 & 25,088 \\
18 & Agent tool call & exec; command: \texttt{\detokenize{python - <<'PY' import os, dpdispatcher root=os.path.dirname(dpdispatcher.__file__) for dirpath, dirs, files in os.walk(root): for f in files: if f.endswith('.py'): p=os.path.join(dirpath,f) with open(p,encoding='utf-8',errors='ignore') as fh: txt=fh.read() if 'submission_args' in txt or 'dargs' in txt: print(p) PY}} & 1,211 & 148 & 25,088 \\
19 & Agent tool call & exec; command: \texttt{\detokenize{python - <<'PY' from dpdispatcher import arginfo import inspect print(inspect.getsource(arginfo)[:8000]) PY}} & 1,142 & 70 & 25,600 \\
20 & Agent tool call & exec; command: \texttt{\detokenize{python - <<'PY' from dpdispatcher.submission import Resources, Task, Submission; from dpdispatcher.machine import Machine; for name, obj in [('machine',Machine.arginfo()),('resources',Resources.arginfo()),('task',Task.arginfo()),('submission',Submission.arginfo())]: print('---',name,'---'); print(obj) PY}} & 753 & 121 & 26,112 \\
21 & Agent tool call & exec; command: \texttt{\detokenize{python - <<'PY' from dpdispatcher.submission import Resources, Task from dpdispatcher.machine import Machine for name, obj in [('machine',Machine.arginfo()),('resources',Resources.arginfo()),('task',Task.arginfo())]: print('\n---',name,'---') print(type(obj)) print(dir(obj)[:20]) print(obj) PY}} & 415 & 128 & 26,624 \\
22 & Agent tool call & exec; command: \texttt{\detokenize{python - <<'PY' from dpdispatcher import Submission import inspect print([m for m in dir(Submission) if 'json' in m or 'argin' in m]) print(inspect.getsource(Submission.submission_from_json)) PY}} & 856 & 93 & 26,624 \\
23 & Agent tool call & exec; command: \texttt{\detokenize{python - <<'PY' from dpdispatcher import Submission import inspect print(inspect.getsource(Submission.deserialize)[:5000]) PY}} & 1,030 & 69 & 26,624 \\
24 & Agent tool call & exec; command: \texttt{\detokenize{python - <<'PY' from dpdispatcher import Submission import inspect src=inspect.getsource(Submission) for line in src.splitlines(): if 'def ' in line or 'task_list' in line or 'belonging_jobs' in line: print(line) PY}} & 824 & 102 & 27,136 \\
25 & Agent tool call & exec; command: \texttt{\detokenize{python - <<'PY' from dpdispatcher import Submission import inspect print(inspect.getsource(Submission.serialize)) print(inspect.getsource(Submission.submission_to_json)) PY}} & 1,559 & 79 & 27,136 \\
26 & Agent tool call & exec; command: \texttt{\detokenize{python - <<'PY' from dpdispatcher import Resources, Machine, Task import inspect print('Resources serialize') print(inspect.getsource(Resources.serialize)) print('Machine deserialize') print(inspect.getsource(Machine.deserialize)) print('Task serialize/deser') print(inspect.getsource(Task.serialize)) print(inspect.getsource(Task.deserialize)) PY}} & 953 & 122 & 28,160 \\
27 & Agent tool call & exec; command: \texttt{\detokenize{python - <<'PY' from dpdispatcher import Machine import inspect print(inspect.getsource(Machine.serialize)) print(inspect.getsource(Machine.load_from_dict)) PY}} & 2,047 & 78 & 27,648 \\
28 & Agent tool call & exec; command: \texttt{\detokenize{python - <<'PY' from dpdispatcher.submission import Job import inspect print(inspect.signature(Job.__init__)) print(inspect.getsource(Job.serialize)) print(inspect.getsource(Job.deserialize)) PY}} & 881 & 86 & 29,184 \\
29 & Agent tool call & exec; command: \texttt{\detokenize{python - <<'PY' from dpdispatcher import Submission import inspect print(inspect.getsource(Submission.__init__)) PY}} & 2,042 & 66 & 28,672 \\
30 & Agent tool call & exec; command: \texttt{\detokenize{python - <<'PY' from dpdispatcher import Submission import inspect print(inspect.getsource(Submission.bind_machine)) print(inspect.getsource(Submission.run_submission)[:4000]) PY}} & 1,293 & 82 & 29,696 \\
31 & Agent tool call & exec; command: \texttt{\detokenize{python - <<'PY' from dpdispatcher import Machine, Resources, Task, Submission; from dpdispatcher.submission import Job; from dpdispatcher.utils.job_status import JobStatus; m=Machine.load_from_dict({...}); r=Resources(number_node=1,cpu_per_node=16,gpu_per_node=0,queue_name='main',...); print(...) PY}} & 1,530 & 234 & 30,208 \\
32 & Agent tool call & exec; command: \texttt{\detokenize{uvx --from openbabel-wheel obabel -V}} & 1,152 & 440 & 31,232 \\
33 & Agent tool call & exec; command: \texttt{\detokenize{python - <<'PY' import inspect import dpdispatcher.run as run print(inspect.getsource(run)[:12000]) PY}} & 363 & 543 & 32,256 \\
34 & Agent tool call & exec; command: \texttt{\detokenize{grep -R "def submit" -n .../site-packages/dpdispatcher /home/ubuntu/miniforge3/bin/dpdisp 2>/dev/null | head -50 && head -200 /home/ubuntu/miniforge3/bin/dpdisp}} & 1,612 & 102 & 32,256 \\
35 & Agent tool call & exec; command: \texttt{\detokenize{python - <<'PY' import dpdispatcher.dpdisp as d, inspect print(inspect.getsource(d)[:12000]) PY}} & 3,366 & 68 & 30,720 \\
36 & Agent tool call & exec; command: \texttt{\detokenize{uvx --from dpdispatcher dpdisp --help && echo '---' && uvx --with dpdispatcher dargs check -h | sed -n '1,120p'}} & 1,212 & 100 & 33,792 \\
37 & Agent tool call & exec; command: \texttt{\detokenize{uvx --from dpdispatcher dpdisp submit --help | sed -n '1,220p'}} & 1,748 & 74 & 33,792 \\
38 & Agent tool call & exec; command: \texttt{\detokenize{python - <<'PY' import pkgutil, dpdispatcher, inspect, os mods=[] for m in pkgutil.walk_packages(dpdispatcher.__path__, dpdispatcher.__name__+'.'): if 'submit' in m.name: mods.append(m.name) print(mods) PY}} & 416 & 103 & 35,328 \\
39 & Agent tool call & exec; command: \texttt{\detokenize{uvx --from dpdispatcher python - <<'PY' import dpdispatcher.dpdisp as d, inspect print(inspect.getsource(d)[:10000]) PY}} & 1,042 & 90 & 34,816 \\
40 & Agent tool call & exec; command: \texttt{\detokenize{uvx --from dpdispatcher python - <<'PY' from dpdispatcher.entrypoints.submit import submit, submission_args import inspect print(inspect.getsource(submission_args)[:7000]) print('--- submit ---') print(inspect.getsource(submit)[:4000]) PY}} & 1,735 & 119 & 35,328 \\
41 & Agent tool call & exec; command: \texttt{\detokenize{uvx --from dpdispatcher python - <<'PY' from dpdispatcher.entrypoints.submit import load_submission_from_json import inspect print(inspect.getsource(load_submission_from_json)[:6000]) PY}} & 2,354 & 98 & 35,328 \\
42 & Agent tool call & update\_plan; steps: \texttt{\detokenize{Inspect methane workspace and m1 config; Resolve molecule identifiers and workflow structure; Write workflow files and scripts; Run lightweight validation}} & 749 & 1,130 & 37,376 \\
43 & Agent tool call & exec; command: \texttt{\detokenize{uvx --from dpdispatcher python - <<'PY' from dpdispatcher.submission import Resources import inspect print(inspect.getsource(Resources.arginfo)[:9000]) PY}} & 859 & 3,873 & 37,376 \\
44 & Agent tool call & exec; command: \texttt{\detokenize{uvx --from dpdispatcher python - <<'PY' from dpdispatcher.machines.slurm import Slurm import inspect print(inspect.getsource(Slurm.resources_arginfo)[:6000]) print('--- script gen snippets ---') for name in ['do_submit','submit_job','gen_script','write_script']: if hasattr(Slurm,name): print(name, inspect.signature(getattr(Slurm,name))) PY}} & 2,819 & 153 & 36,864 \\
45 & Agent tool call & exec; command: \texttt{\detokenize{uvx --from dpdispatcher python - <<'PY' from dpdispatcher.machines.slurm import Slurm import inspect print(inspect.getsource(Slurm.resources_subfields)[:9000]) print(inspect.getsource(Slurm.gen_script)[:10000]) PY}} & 2,078 & 116 & 37,888 \\
46 & Agent tool call & exec; command: \texttt{\detokenize{uvx --from dpdispatcher python - <<'PY' from dpdispatcher.machines import slurm import inspect print(inspect.getsource(slurm)[:20000]) PY}} & 831 & 93 & 39,424 \\
47 & Agent tool call & exec; command: \texttt{\detokenize{uvx --from dpdispatcher python - <<'PY' from dpdispatcher.machine import Machine import inspect print(inspect.getsource(Machine.gen_script)[:10000]) print('--- header ---') print(inspect.getsource(Machine.gen_script_header)[:5000]) print('--- command ---') print(inspect.getsource(Machine.gen_script_command)[:8000]) print('--- env ---') print(inspect.getsource(Machine.gen_command_env_cuda_...}} & 4,214 & 170 & 39,424 \\
48 & Agent tool call & exec; command: \texttt{\detokenize{uvx --from dpdispatcher python - <<'PY' from dpdispatcher import machine print(machine.script_template) print('--- command template ---') print(machine.script_command_template) print('--- end ---') print(machine.script_end_template) PY}} & 921 & 110 & 43,520 \\
49 & Agent tool call & exec; command: \texttt{\detokenize{uvx --from dpdispatcher python - <<'PY' from dpdispatcher.machine import Machine; import inspect; print(inspect.getsource(Machine.gen_script_env)[:7000]); print(inspect.getsource(Machine.gen_script_custom_flags_lines)[:3000]); print(inspect.getsource(Machine.gen_script_run_command)[:4000]); print(inspect.getsource(Machine.gen_script_end)[:4000]) PY}} & 738 & 148 & 44,032 \\
50 & Agent tool call & exec; command: \texttt{\detokenize{uvx --from dpdispatcher python - <<'PY' from dpdispatcher import machine print(machine.script_env_template) PY}} & 1,030 & 80 & 44,544 \\
51 & Agent tool call & read; path: \texttt{\detokenize{skills/packmol-generate-mixture/SKILL.md}}\newline read; path: \texttt{\detokenize{skills/lammps-deepmd/SKILL.md}}\newline read; path: \texttt{\detokenize{skills/dpdata-cli/SKILL.md}} & 671 & 665 & 45,056 \\
52 & Agent tool call & exec; command: \texttt{\detokenize{set -euo pipefail; WF=/home/ubuntu/.openclaw/workspace/methane/ch4_o2_combustion_workflow; rm -rf "$WF"; mkdir -p "$WF"/{inputs,gaussian/ch4,gaussian/o2,optimized,packing,model,md,analysis,submissions,scripts,logs,metadata}; cat > "$WF/metadata/smiles_sources.tsv" <<'EOF' ...; uvx --from openbabel-wheel obabel -:'C' --gen3d -oxyz -O "$WF/inputs/ch4_initial.xyz"; uvx --from openbabel-wheel obabel -:'O=O' --gen3d -oxyz -O "$WF/inputs/o2_initial.xyz"; find "$WF" -maxdepth 2 -type f | sort}} & 6,524 & 2,209 & 45,568 \\
53 & Agent tool call (terminated) & exec; stopReason: error; errorMessage: terminated; command: \texttt{\detokenize{cat > /tmp/create_workflow.py <<'PY' from pathlib import Path import json, math, textwrap base = Path('$WF') workspace = Path('/home/ubuntu/.openclaw/workspace/methane') # Helpers masses = ( 'H': 1.00784, 'C': 12.011, 'O': 15.999, ) def read_xyz(path): lines = Path(path).read_text().splitlines() n = int(lines[0].strip()) atoms = [] for line in lines[2:2...}} & 0 & 0 & 0 \\
54 & Agent tool call & exec; command: \texttt{\detokenize{cat $WF/inputs/*.xyz}} & 376 & 64 & 52,224 \\
55 & Agent tool call & exec; command: \texttt{\detokenize{set -euo pipefail; WF=/home/ubuntu/.openclaw/workspace/methane/ch4_o2_combustion_workflow; curl -L --fail --retry 3 --retry-delay 2 -o "$WF/model/graph.pb" "https://github.com/tongzhugroup/Chapter13-tutorial/raw/master/graph.pb"; cp "$WF/model/graph.pb" "$WF/md/graph.pb"; ls -lh "$WF/model/graph.pb" "$WF/md/graph.pb"}} & 593 & 6,868 & 52,224 \\
56 & Agent tool call & exec; command: \texttt{\detokenize{set -euo pipefail WF=$WF for URL in \ 'https://raw.githubusercontent.com/tongzhugroup/Chapter13-tutorial/master/graph.pb' \ 'https://cdn.jsdelivr.net/gh/tongzhugroup/Chapter13-tutorial@master/graph.pb'; do echo "Trying $URL" if curl -L --fail --connect-timeout 15 --max-time 60 -o "$WF/model/graph.pb.tmp" "$URL"; then mv "$WF/model/graph.pb.tmp" "$WF/mod...}} & 442 & 371 & 59,904 \\
57 & Agent tool call & web\_fetch; url: \traceurl{https://raw.githubusercontent.com/tongzhugroup/Chapter13-tutorial/master/input.lammps} & 1,292 & 2,166 & 64,000 \\
58 & Agent tool call & exec; command: \texttt{\detokenize{uvx --from dpdispatcher python - <<'PY' from dpdispatcher.machine import Machine import inspect print(inspect.getsource(Machine.gen_script_wait)) PY}} & 916 & 911 & 65,024 \\
59 & Agent tool call & exec; command: \texttt{\detokenize{cat > /tmp/write_workflow.py <<'PY' from pathlib import Path import json, math, textwrap, os, shutil WF = Path('/home/ubuntu/.openclaw/workspace/methane/ch4_o2_combustion_workflow') ... PY; python /tmp/write_workflow.py; find $WF -maxdepth 3 -type f | sort}} & 661 & 8,062 & 65,536 \\
60 & Agent tool call & exec; command: \texttt{\detokenize{cat > /tmp/write_workflow.py <<'PYWF' from pathlib import Path import json WF = Path('/home/ubuntu/.openclaw/workspace/methane/ch4_o2_combustion_workflow') WF.mkdir(parents=True, exist_ok=True) ... PYWF; python /tmp/write_workflow.py; find $WF -maxdepth 3 -type f | sort}} & 8,268 & 8,712 & 66,048 \\
61 & Agent tool call & exec; command: \texttt{\detokenize{python - <<'PY' from pathlib import Path WF=Path('$WF') extract=(WF/'scripts'/'extract_gaussian_geometry.py').read_text() for species in ['ch4','o2']: (WF/'gaussian'/species/'extract_gaussian_geometry.py').write_text(extract) (WF/'gaussian'/species/'extract_gaussian_geometry.py').chmod(0o755) run=WF/'gaussian'/species/'run_gaussian.sh' txt=run.read_text...}} & 8,995 & 901 & 74,240 \\
62 & Agent tool call & update\_plan; steps: \texttt{\detokenize{Inspect methane workspace and m1 config; Resolve molecule identifiers and workflow structure; Write workflow files and scripts; Run lightweight validation}} & 873 & 96 & 82,944 \\
63 & Agent tool call & exec; command: \texttt{\detokenize{cd $WF && set -e uvx --with dpdispatcher dargs check --allow-ref -f dpdispatcher.entrypoints.submit.submission_args submissions/gaussian_opt_freq.json uvx --with dpdispatcher dargs check --allow-ref -f dpdispatcher.entrypoints.submit.submission_args submissions/deepmd_nvt.json python -m py_compile...}} & 473 & 265 & 83,456 \\
64 & Agent tool call & exec; command: \texttt{\detokenize{cd $WF && rm -rf logs/packmol_smoke && python scripts/build_reaction_system.py --ch4-xyz inputs/ch4_initial.xyz --o2-xyz inputs/o2_initial.xyz --outdir logs/packmol_smoke --density 0.25 --n-ch4 1 --n-o2 1 --tolerance 2.0 > logs/packmol_smoke.json && python - <<'PY' from pathlib import Path import json summary=json.loads(Path('logs/packmol_smoke/system_s...}} & 409 & 693 & 83,968 \\
65 & Agent tool call & exec; command: \texttt{\detokenize{cd $WF && sed -n '1,120p' gaussian/ch4/ch4.gjf && echo '--- O2' && sed -n '1,120p' gaussian/o2/o2.gjf && echo '--- LAMMPS' && sed -n '1,180p' md/input.lammps && echo '--- submissions refs' && grep -R 'machine_m1\|mem\|gpu_per_node\|cpu_per_node\|queue_name\|source' -n submissions md/run_md.sh gaussian/*/run_gaussian.sh README.md | sed -n '1,200p'}} & 281 & 187 & 84,480 \\
66 & Agent tool call & update\_plan; steps: \texttt{\detokenize{Inspect methane workspace and m1 config; Resolve molecule identifiers and workflow structure; Write workflow files and scripts; Run lightweight validation}} & 1,507 & 611 & 84,480 \\
67 & \textbf{Agent message} & Done — I created the complete workflow here: /home/ubuntu/.openclaw/workspace/methane/ch4\_o2\_combustion\_workflow Key files: - README.md — full workflow instructions - ch4\_o2\_combustion\_workflow.json — workflow manifest - inputs/ch4\_initial.xyz, inputs/o2\_initial.xyz — Open Babel 3D coordinates - gaussian/ch4/ch4.gjf, gaussian/o2/o2.gjf — B3LYP/6-31G(d,p) opt+freq Gaussian inputs - submissions/gaussian\_opt\_freq.json — dpdispatcher Gaussian submission using ../machine\_m1.json - scripts/02\_build\_system.sh + scripts/build\_reaction\_system.py — packs 50 CH₄ + 100 O₂ at 0.25 g/cm³ - md/input.lammps — DeePMD/LAMMPS NVT 3000 K, 1 ns, 0.1 fs, units real, dump every 1000 steps - submissions/deepmd\_nvt.json — dpdispatcher MD submission, Slurm main, 1 GPU - analysis/analyze\_reaction\_pathways.py — first-pass reaction pathway/species-count analysis - launch\_tmux.sh — detached launcher using tmux I also downloaded the DeePMD potential to: - model/graph.pb - md/graph.pb Validation completed: - dpdispatcher schema checks passed for both Gaussian and MD submission JSONs. - Python helper scripts compile cleanly. - Packmol/data-generation smoke test passed. - Confirmed generated SMILES/coordinates: - methane: C - oxygen: O=O I did not submit the long Gaussian/MD jobs yet. To launch the full workflow from the workflow directory: \texttt{\detokenize{./launch_tmux.sh}} or stepwise: \texttt{\detokenize{./scripts/01_submit_gaussian.sh / ./scripts/02_build_system.sh / ./scripts/03_submit_md.sh / ./analysis/run_analysis.sh}} & 592 & 974 & 85,504 \\
68 & \textbf{User message} & Thanks! Please go ahead and use dpdispatcher to submit jobs, use reacnetgenerator to analyze the trajectory, and report the reactions in the trajectory. &  &  &  \\
69 & Agent tool call & read; path: \texttt{\detokenize{skills/reacnetgenerator/SKILL.md}} & 70,788 & 493 & 15,872 \\
70 & \textbf{Agent message + tool call} & I'll submit this as a detached tmux-managed workflow, but first I'm replacing the placeholder analysis with ReacNetGenerator so the trajectory report is generated automatically after MD completes.\newline update\_plan; steps: \texttt{\detokenize{Inspect workflow state and ReacNetGenerator CLI; Add ReacNetGenerator analysis step; Submit workflow via tmux/dpdispatcher; Verify submission and set monitoring}} & 1,775 & 1,052 & 86,528 \\
71 & Agent tool call & exec; command: \texttt{\detokenize{find $WF -maxdepth 3 -type f | sort | sed -n '1,220p'}}\newline exec; command: \texttt{\detokenize{uvx --from reacnet-md-tools rng-pipeline --help | sed -n '1,220p'}}\newline exec; command: \texttt{\detokenize{uvx --from reacnet-md-tools rng-query --help | sed -n '1,220p'}} & 374 & 257 & 88,064 \\
72 & Agent tool call & process & 1,922 & 62 & 88,064 \\
73 & Agent tool call & process & 848 & 61 & 89,600 \\
74 & Agent tool call & exec; command: \texttt{\detokenize{uvx --from reacnet-md-tools rng-query species --help | sed -n '1,200p'}}\newline exec; command: \texttt{\detokenize{uvx --from reacnet-md-tools rng-query rxn-formula --help | sed -n '1,220p'}}\newline exec; command: \texttt{\detokenize{uvx --from reacnet-md-tools rng-query next --help | sed -n '1,220p'}} & 623 & 638 & 90,112 \\
75 & Agent tool call & exec; command: \texttt{\detokenize{cd $WF && python - <<'PY' from pathlib import Path import json WF=Path('.') run_analysis = r'''#!/usr/bin/env bash set -euo pipefail ...''' ... PY}} & 1,182 & 2,685 & 90,624 \\
76 & Agent tool call & process & 2,886 & 102 & 91,648 \\
77 & Agent tool call & exec; command: \texttt{\detokenize{cd $WF && python - <<'PY' from pathlib import Path import json for name in ['gaussian_opt_freq.json','deepmd_nvt.json']: p=Path('submissions')/name j=json.loads(p.read_text()) j['machine']=('$ref':'../../machine_m1.json') p.write_text(json.dumps(j, indent=2)+'\n') # dargs resolves refs relative to cwd, so validate from submissions/; dpdisp itself also r...}} & 498 & 3,491 & 94,208 \\
78 & Agent tool call & process & 1,098 & 62 & 94,208 \\
79 & Agent tool call & exec; command: \texttt{\detokenize{cd $WF && sed -n '1,120p' launch_tmux.sh && echo --- && sed -n '1,100p' scripts/01_submit_gaussian.sh && echo --- && sed -n '1,120p' analysis/run_reacnet.sh && echo --- && grep -R 'machine_m1' -n README.md submissions/*.json}} & 156 & 141 & 95,232 \\
80 & Agent tool call & exec; command: \texttt{\detokenize{cd $WF && tmux capture-pane -t ch4_o2_combustion -p -S -200 | tail -120 && echo '--- local gaussian files' && find gaussian -maxdepth 2 -type f \( -name '*.log' -o -name '*_opt.xyz' -o -name 'log' -o -name 'err' \) -printf '\%p \%s bytes\n' | sort && echo '--- newest logs' && ls -lt logs | head}} & 724 & 406 & 95,744 \\
81 & Agent tool call & exec; command: \texttt{\detokenize{cd $WF && tmux capture-pane -t ch4_o2_combustion -p -S -200 | tail -120 && echo '--- dpdispatcher log tail' && tail -80 dpdispatcher.log 2>/dev/null || true}} & 725 & 1,670 & 96,256 \\
82 & Agent tool call & cron; action: \texttt{\detokenize{add}}; job: \texttt{\detokenize{Check CH4/O2 combustion workflow progress}} & 1,617 & 219 & 96,768 \\
83 & Agent tool call & update\_plan; steps: \texttt{\detokenize{Inspect workflow state and ReacNetGenerator CLI; Add ReacNetGenerator analysis step; Submit workflow via tmux/dpdispatcher; Verify submission and set monitoring}} & 722 & 107 & 98,304 \\
84 & \textbf{Agent message} & Submitted and monitoring. Current status: - Gaussian B3LYP/6-31G(d,p) opt/freq jobs submitted via dpdispatcher on m1 and already finished: - Slurm jobs: 2563, 2564 - Optimized geometries were extracted and the 50 CH₄ + 100 O₂ box was built: - density: 0.25 g/cm³ - box length: 29.844352867 Å - atoms: C=50, H=200, O=200, total 450 - DeePMD/LAMMPS MD was submitted via dpdispatcher: - Slurm job: 2565 - remote root: /home/jzzeng/dpgen\_workdir/9b5bf50cc6099c2ed0e5b45949405de233b42a9a - local workflow: /home/ubuntu/.openclaw/workspace/methane/ch4\_o2\_combustion\_workflow I also replaced the trajectory analysis step with ReacNetGenerator: - Script: analysis/run\_reacnet.sh - Output target: analysis/reacnet/ - Final report target: analysis/reacnet/reaction\_report.md I set a 30-minute monitor. When the MD finishes and ReacNetGenerator completes, I’ll report the reactions found in the trajectory. & 333 & 297 & 98,816 \\
85 & Agent tool call & exec; command: \texttt{\detokenize{cd $WF && ( tmux has-session -t ch4_o2_combustion 2>/dev/null && echo TMUX=running && tmux capture-pane -t ch4_o2_combustion -p -S -120 | tail -80 || echo TMUX=not_running; ) && echo '--- dpdispatcher tail ---' && tail -60 dpdispatcher.log 2>/dev/null || true && echo '--- md files ---' && find md -maxdepth 1 -type f -printf '\%f \%s bytes \%TY-\%Tm-\%Td \%TH:...}} & 83,904 & 550 & 11,776 \\
86 & \textbf{Agent message + tool call} & MD is failing immediately on m1, so I'm going to diagnose the remote job logs now rather than waiting silently.\newline exec; command: \texttt{\detokenize{cd $WF && echo '--- workflow logs ---' && for f in logs/workflow_*.log; do echo '###' $f; tail -160 "$f"; done && echo '--- dpdispatcher record dirs/files ---' && find . -maxdepth 3 -type f \( -name '*.json' -o -name 'log' -o -name 'err' \) -printf '\%p \%s bytes \%TY-\%Tm-\%Td \%TH:\%TM\n' | sort | tail -100 && echo '--- dpdispatcher records home ---' && find...}} & 1,976 & 446 & 95,232 \\
87 & \textbf{Agent message + tool call} & The MD job failed quickly because the DeePMD conda activation script is incompatible with \texttt{\detokenize{set -u}} in \texttt{\detokenize{run_md.sh}} (\texttt{\detokenize{ADDR2LINE: unbound variable}}). I found the cause; I'm patching that wrapper and resubmitting the MD stage now.\newline exec; command: \texttt{\detokenize{cd $WF && python - <<'PY' from pathlib import Path p=Path('md/run_md.sh') t=p.read_text() old='''#!/usr/bin/env bash set -euo pipefail source /group/software/deepmd-kit-3.1.1/bin/activate /group/software/deepmd-kit-3.1.1 export OMP_NUM_THREADS=$(OMP_NUM_THREADS:-1) ''' new='''#!/usr/bin/env bash set -eo pipefail # The DeePMD conda activation scripts ref...}} & 3,287 & 1,088 & 96,768 \\
88 & Agent tool call & process & 2,466 & 60 & 100,864 \\
89 & Agent tool call & exec; command: \texttt{\detokenize{cd $WF && ( tmux has-session -t ch4_o2_md_resume 2>/dev/null && echo TMUX=running && tmux capture-pane -t ch4_o2_md_resume -p -S -160 | tail -120 || echo TMUX=not_running; ) && echo '--- dpdispatcher tail ---' && tail -100 dpdispatcher.log && echo '--- md local outputs ---' && find md -maxdepth 1 -type f -printf '\%f \%s bytes \%TY-\%Tm-\%Td \%TH:\%TM\n' | sort}} & 869 & 554 & 102,912 \\
90 & Agent tool call & exec; command: \texttt{\detokenize{cd $WF && ( tmux has-session -t ch4_o2_md_resume 2>/dev/null && echo TMUX=running && tmux capture-pane -t ch4_o2_md_resume -p -S -160 | tail -120 || echo TMUX=not_running; ) && echo '--- dpdispatcher tail ---' && tail -80 dpdispatcher.log && echo '--- workflow resume logs ---' && for f in logs/md_reacnet_resume_*.log; do [ -e "$f" ] && ( echo "### $f";...}} & 2,691 & 531 & 103,424 \\
91 & \textbf{Agent message} & MD needed intervention: it failed immediately because the DeePMD activation script tripped over \texttt{\detokenize{set -u}} (\texttt{\detokenize{ADDR2LINE: unbound variable}}). I patched \texttt{\detokenize{md/run_md.sh}}, reset the dpdispatcher fail count, and resumed the MD stage. Current resumed MD job: - Slurm job: \texttt{\detokenize{2569}} - Status at last check: waiting - Resume session: \texttt{\detokenize{ch4_o2_md_resume}} I'll keep watching it and only report again when it progresses, fails with a new issue, or finishes and ReacNetGenerator produces the reaction report. & 2,403 & 639 & 105,984 \\
92 & Agent tool call & exec; command: \texttt{\detokenize{cd $WF && echo '--- local processes ---' && ps -ef | grep -E 'dpdisp|ch4_o2|tmux|uvx|ssh' | grep -v grep && echo '--- tmux pane processes ---' && tmux list-panes -a -F '#S:#I.#P pid=#(pane_pid) cmd=#(pane_current_command)' 2>/dev/null || true && echo '--- latest dpdispatcher record mtime ---' && ls -l /home/ubuntu/.dpdispatcher/submission/9b5bf50cc6099c...}} & 2,679 & 459 & 115,200 \\
93 & Agent tool call & read; path: \texttt{\detokenize{skills/reacnetgenerator/SKILL.md}} & 12,522 & 116 & 0 \\
94 & Agent tool call & read; path: \texttt{\detokenize{methane/ch4_o2_combustion_workflow/md/run_md.sh}} & 3,585 & 195 & 16,896 \\
95 & Agent tool call & read; path: \texttt{\detokenize{skills/dpdisp-submit/SKILL.md}} & 835 & 195 & 19,968 \\
96 & \textbf{Agent message} & The CH4/O2 workflow is blocked at the DeePMD/LAMMPS MD stage. What completed: - Gaussian CH4 and O2 jobs finished successfully. - Optimized geometries were retrieved: - gaussian/ch4/ch4\_opt.xyz - gaussian/o2/o2\_opt.xyz - Packing/system build completed: - 50 CH4 + 100 O2 - 450 atoms total - density 0.25 g/cm³ - LAMMPS data: md/data.ch4\_o2 What failed: - The DeePMD MD job was submitted via dpdispatcher, but Slurm job 111f483fbb2c788bb4b54698c9c97c76fcdfbba3 repeatedly terminated and failed after multiple resubmissions. - No MD trajectory was retrieved. - Therefore ReacNetGenerator analysis has not run, and analysis/reacnet/reaction\_report.md does not exist yet. Relevant blocker from logs/workflow\_20260503\_003701.log / dpdispatcher.log: \texttt{\detokenize{RuntimeError: job 111f483fbb2c788bb4b54698c9c97c76fcdfbba3 2568 failed 4 times. / Possible remote error message: / /group/software/deepmd-kit-3.1.1/etc/conda/activate.d/activate-binutils_linux-64.sh: line 68: ADDR2LINE: unbound variable}} Likely cause: - The remote DeePMD-kit activation script is tripping over an unset ADDR2LINE variable during environment activation. - Until that activation issue is fixed, LAMMPS/DeePMD cannot start, so there is no trajectory for ReacNetGenerator and no reactions to report yet. & 2,862 & 553 & 20,480 \\
97 & Agent tool call & read; path: \texttt{\detokenize{openclaw/skills/tmux/SKILL.md}} & 12,522 & 121 & 0 \\
98 & Agent tool call & update\_plan; steps: \texttt{\detokenize{Inspect workflow files/logs; Check tmux/job status; Summarize or stay quiet}} & 1,314 & 56 & 12,288 \\
99 & Agent tool call & read; path: \texttt{\detokenize{methane/ch4_o2_combustion_workflow/md/run_md.sh}} & 3,676 & 211 & 15,872 \\
100 & Agent tool call & read; path: \texttt{\detokenize{skills/dpdisp-submit/SKILL.md}} & 414 & 383 & 19,456 \\
101 & Agent tool call & update\_plan; steps: \texttt{\detokenize{Inspect workflow files/logs; Check tmux/job status; Summarize or stay quiet}} & 3,085 & 280 & 19,456 \\
102 & \textbf{Agent message} & The CH4/O2 combustion workflow is blocked. What I found: - tmux session ch4\_o2\_combustion is not running. - Gaussian jobs completed successfully: - CH4 job 2563 finished - O2 job 2564 finished - The packed system was built: 50 CH4 + 100 O2, 450 atoms total. - The DeePMD/LAMMPS MD job repeatedly failed and was resubmitted until DPDispatcher stopped after 4 failures. - analysis/reacnet/reaction\_report.md is missing, so ReacNetGenerator analysis has not completed and there are no reactions to report yet. Relevant blocker from logs/workflow\_20260503\_003701.log / dpdispatcher.log: \texttt{\detokenize{RuntimeError: job 111f483fbb2c788bb4b54698c9c97c76fcdfbba3 2568 failed 4 times. / Possible remote error message: / /group/software/deepmd-kit-3.1.1/etc/conda/activate.d/activate-binutils_linux-64.sh: line 68: ADDR2LINE: unbound variable}} Likely cause: the remote DeePMD activation script is incompatible with set -u / nounset behavior during environment activation. The local md/run\_md.sh already avoids enabling set -u until after activation, but the submitted job still hit ADDR2LINE: unbound variable, possibly from an older forwarded script or activation shell behavior. Next useful step would be to patch md/run\_md.sh more defensively, e.g. predefine the binutils variables or avoid set -u entirely for the MD run, then resubmit the DPDispatcher MD job. & 622 & 378 & 22,016 \\
103 & Agent tool call & exec; command: \texttt{\detokenize{tmux has-session -t ch4_o2_combustion 2>/dev/null && ( echo '=== tmux pane ==='; tmux capture-pane -t ch4_o2_combustion -p -S - | tail -80; ) || echo 'NO_TMUX_SESSION ch4_o2_combustion'}}\newline exec; command: \texttt{\detokenize{cd $WF 2>/dev/null || exit 0 printf '=== latest tracked files ===\n' find . -maxdepth 4 -type f \( -name 'dpdispatcher.log' -o -name 'reaction_report.md' -o -name 'log.lammps' -o -name '*.lammpstrj' -o -name 'err' -o -name 'log' -o -name '*.final' \) -printf '\%TY-\%Tm-\%Td \%TH:\%TM:\%TS \%p \%s bytes\n' | sort | tail -100 printf '\n=== dpdispatcher tail ===\n...}} & 575 & 444 & 51,712 \\
104 & Agent tool call & exec; command: \texttt{\detokenize{cd $WF printf '=== reaction ===\n'; cat md/ch4_o2_3000K.lammpstrj.reaction printf '\n=== species ===\n'; cat md/ch4_o2_3000K.lammpstrj.species printf '\n=== table ===\n'; cat md/ch4_o2_3000K.lammpstrj.table printf '\n=== moname head ===\n'; head -80 md/ch4_o2_3000K.lammpstrj.moname}}\newline exec; command: \texttt{\detokenize{cd $WF printf '=== analysis scripts ===\n'; sed -n '1,220p' analysis/report_reacnet_reactions.py printf '\n=== run_reacnet ===\n'; sed -n '1,180p' analysis/run_reacnet.sh}} & 3,882 & 352 & 51,712 \\
105 & \textbf{Agent message} & The workflow has completed enough to report reactions. Status: - Gaussian CH4/O2 optimization jobs finished. - DeePMD/LAMMPS MD finished via DPDispatcher: job 2569 finished at 2026-05-04 09:57. - MD trajectory exists: md/ch4\_o2\_3000K.lammpstrj \textasciitilde{}135 MB. - ReacNetGenerator outputs were generated in md/, including: - md/ch4\_o2\_3000K.lammpstrj.reaction - md/ch4\_o2\_3000K.lammpstrj.species - md/ch4\_o2\_3000K.lammpstrj.table - Caveat: analysis/reacnet/reaction\_report.md exists but says reaction/species files “not found” because the report script looked under analysis/reacnet, while the actual ReacNetGenerator output landed in md/. Main reaction/species evolution from ReacNetGenerator: - Initial system: 50 CH4 + 100 O2 - Final sampled timestep 10000000: - CH4: 8 remaining - O2: 15 remaining - H2O: 83 - CO: 22 - CO2: 16 - O–O / peroxide-like species: 14 - minor OCO / C2 oxygenated intermediates also present Top reported reaction channels by count: 1. CH4 -> H2O count 128 2. O2 -> O–O count 122 3. O–O -> O2 count 89 4. O2 -> H2O count 50 5. O–O -> H2O count 31 6. HOO -> H2O count 22 7. H2 -> H2O count 20 8. CH4 -> CH3 count 20 9. CH3 -> H2O count 18 10. CH4 -> H2 count 17 11. formaldehyde-like CH2O -> H2O count 16 12. CH4 -> CO count 13 13. OCO -> CO2 count 10 14. O2 -> CO count 10 15. CO -> CO2 count 9 Overall interpretation: the 3000 K trajectory shows substantial methane oxidation. CH4 is consumed from 50 to 8 molecules, O2 from 100 to 15, with H2O as the dominant product and significant CO/CO2 formation. The network includes expected combustion intermediates/radicals such as CH3, CH2O/formaldehyde-like species, HOO/HO2, H2O2-like species, CO, OCO, and CO2. & 10,340 & 698 & 55,296 \\
106 & \textbf{User message} & ReacNetGenerator generates analyzed results under the same directory as the trajectory file. &  &  &  \\
107 & Agent tool call & read; path: \texttt{\detokenize{skills/reacnetgenerator/SKILL.md}} & 75,539 & 138 & 0 \\
108 & Agent tool call & edit & 1,917 & 676 & 75,264 \\
109 & Agent tool call & exec; command: \texttt{\detokenize{cd $WF python3 analysis/report_reacnet_reactions.py md > analysis/reacnet/reaction_report.md head -80 analysis/reacnet/reaction_report.md}} & 587 & 90 & 76,800 \\
110 & Agent tool call & edit & 3,523 & 675 & 77,312 \\
111 & Agent tool call & exec; command: \texttt{\detokenize{cd $WF python3 analysis/report_reacnet_reactions.py md > analysis/reacnet/reaction_report.md head -70 analysis/reacnet/reaction_report.md}} & 658 & 90 & 80,384 \\
112 & Agent tool call & edit & 1,773 & 572 & 80,896 \\
113 & Agent tool call & exec; command: \texttt{\detokenize{cd $WF python3 analysis/report_reacnet_reactions.py md > analysis/reacnet/reaction_report.md head -55 analysis/reacnet/reaction_report.md}} & 859 & 90 & 82,432 \\
114 & \textbf{Agent message} & You're right --- thanks for catching that. I updated the workflow/reporting so it now looks in the trajectory directory (\texttt{\detokenize{md/}}) instead of \texttt{\detokenize{analysis/reacnet}}. I also regenerated \texttt{\detokenize{analysis/reacnet/reaction_report.md}}; it now correctly points to: - \texttt{\detokenize{md/ch4_o2_3000K.lammpstrj.reaction}} - \texttt{\detokenize{md/ch4_o2_3000K.lammpstrj.species}} and parses the reaction counts properly. The top channel is now reported as: \texttt{\detokenize{[H][C]([H])([H])[H] -> [H][O][H] weight 128 }} instead of treating \texttt{\detokenize{128}} as part of the reaction string. & 1,674 & 155 & 82,944 \\
115 & \textbf{User message} & Please describe what reactions happen in this simulation. &  &  &  \\
116 & \textbf{Agent message} & The simulation shows methane oxidation/combustion at 3000 K. Main picture: - Initial system: 50 CH₄ + 100 O₂ - By the final sampled frame: - CH₄ drops to 8 - O₂ drops to 15 - H₂O rises to 83 - CO reaches 22 - CO₂ reaches 16 So the dominant chemistry is: \texttt{\detokenize{CH4 + O2 → H2O + CO / CO2 + intermediates}} Key reaction families observed by ReacNetGenerator: 1. Methane activation / H abstraction - CH₄ forms methyl-like fragments: \texttt{\detokenize{CH4 → CH3 / CH4 → H2 / CH4 → H}} This is the start of methane oxidation. 2. Water formation is dominant Many channels lead to H₂O: \texttt{\detokenize{CH4 → H2O / O2 / O–O species → H2O / HO2 → H2O / H2 → H2O / CH3 → H2O / CH2O → H2O}} The top ReacNet channel was: \texttt{\detokenize{CH4 → H2O    count 128}} 3. Oxygen bond rearrangement / activated oxygen O₂ repeatedly converts between double-bond and single-bond/peroxide-like forms: \texttt{\detokenize{O2 → O–O / O–O → O2 / O2 → HO2 / O–O → HO2}} This indicates active O–O bond breaking/reforming and radical/peroxide chemistry. 4. Partial oxidation to CO Methane and oxygenated C/H/O intermediates form CO: \texttt{\detokenize{CH4 → CO / CH2O → CO / O2 → CO / O–O → CO}} CO becomes a major product/intermediate. 5. Further oxidation to CO₂ CO and OCO-like intermediates oxidize to CO₂: \texttt{\detokenize{CO → CO2 / OCO → CO2 / O2 → CO2 / H2O → CO2}} Final CO₂ count is significant, but CO remains higher than CO₂, so oxidation is incomplete within this trajectory. 6. Detected intermediates ReacNetGenerator also sees transient/low-count species such as: - CH₃ methyl radical-like species - CH₂O formaldehyde-like species - CH₃OH / CH₂OH-like oxygenated hydrocarbons - HO₂ - H₂O₂-like species - H₂ - CO - CO₂ - OCO / carbonate-like oxygenated carbon fragments Overall: this is a reactive methane combustion trajectory where CH₄ is mostly consumed, O₂ is strongly depleted, water is the dominant product, and carbon is split between CO and CO₂. The mechanism passes through methyl, formaldehyde-like, HO₂/peroxide, and CO intermediates before forming CO₂. & 3,909 & 950 & 80,896 \\
117 & \textbf{User message} & Please describe what reactions happen in this simulation. &  &  &  \\
118 & Agent tool call & read; path: \texttt{\detokenize{methane/ch4_o2_combustion_workflow/analysis/reacnet/reaction_report.md}}\newline read; path: \texttt{\detokenize{methane/ch4_o2_combustion_workflow/md/ch4_o2_3000K.lammpstrj.species}} & 1,638 & 325 & 89,088 \\
119 & \textbf{Agent message} & Based on the ReacNetGenerator outputs, this trajectory shows methane oxidation/combustion at 3000 K. Initial composition: \texttt{\detokenize{50 CH4 + 100 O2}} Final sampled composition: \texttt{\detokenize{CH4: 8 / O2: 15 / H2O: 83 / CO: 22 / CO2: 16}} So the main outcome is: \texttt{\detokenize{CH4 + O2 → mostly H2O + CO + CO2}} with incomplete oxidation because CO remains abundant. Key reaction behavior: 1. Methane is activated and consumed - CH4 decreases from 50 to 8. - ReacNet detects methyl / dehydrogenated fragments: \texttt{\detokenize{CH4 → CH3 / CH4 → H2 / CH4 → H / CH4 → CH2O / CH4 → CO}} 2. Water is the dominant product - H2O grows to 83 molecules. - Major water-forming channels include: \texttt{\detokenize{CH4 → H2O / O2 / O–O species → H2O / HO2 → H2O / H2 → H2O / CH3 → H2O / CH2O → H2O}} 3. Oxygen undergoes strong bond rearrangement - O2 is depleted from 100 to 15. - ReacNet sees frequent interconversion: \texttt{\detokenize{O2 ⇌ O–O/peroxide-like species / O2 → HO2 / O–O → HO2 / H2O}} 4. Carbon goes through partial oxidation - Formaldehyde-like and methanol-like intermediates appear: \texttt{\detokenize{CH3 / CH2O / CH2OH / CH3O-like species / CH3OH-like species}} - These then form CO: \texttt{\detokenize{CH4 → CO / CH2O → CO / oxygenated C/H/O species → CO}} 5. CO is further oxidized to CO2 - CO2 forms, but not all CO is consumed: \texttt{\detokenize{CO → CO2 / OCO-like species ⇌ CO2 / O2 + carbon fragments → CO2}} Top high-count ReacNet channels include: \texttt{\detokenize{CH4 → H2O        128 / O2 → O–O         122 / O–O → O2          89 / O2 → H2O          50 / O–O → H2O         31 / HO2 → H2O         22 / H2 → H2O          20 / CH4 → CH3         20 / CH3 → H2O         18 / CH4 → H2          17 / CH2O → H2O        16 / CH4 → CO          13 / CO → CO2           9}} One caveat: ReacNetGenerator reaction lines are graph/species transitions, not necessarily isolated elementary balanced reactions. But the net chemistry is clear: CH4 is oxidized through radical/peroxide and oxygenated hydrocarbon intermediates, producing mostly H2O, with substantial CO and some CO2. & 3,017 & 1,241 & 90,624 \\
120 & \textbf{User message} & What are key intermediates in this reaction? &  &  &  \\
121 & \textbf{Agent message} & Key intermediates detected by ReacNetGenerator in this CH₄/O₂ combustion trajectory are: 1. CH₃ / methyl-like radicals \texttt{\detokenize{[H][C]([H])[H]}} Formed from methane C–H activation: \texttt{\detokenize{CH4 → CH3}} 2. CH₂O / formaldehyde-like species \texttt{\detokenize{[H][C]([H])=[O]}} This is an important oxygenated carbon intermediate on the path from CH₄ toward CO. 3. CH₃O / CH₂OH / methoxy–hydroxymethyl-like species Examples in the ReacNet output: \texttt{\detokenize{[H][C]([H])[O] / [H][O][C]([H])([H])[H]}} These are transient oxygenated hydrocarbon fragments. 4. HO₂ / hydroperoxyl-like species \texttt{\detokenize{[H][O][O]}} This appears repeatedly and converts to water or back to O₂/O–O species: \texttt{\detokenize{HO2 → H2O / HO2 → O2}} 5. H₂O₂ / hydrogen peroxide-like species \texttt{\detokenize{[H][O][O][H]}} Low abundance, but present as part of peroxide chemistry. 6. O–O / peroxide-like oxygen species \texttt{\detokenize{[O][O]}} Very active intermediate, with strong interconversion: \texttt{\detokenize{O2 ⇌ O–O / O–O → H2O / HO2 / CO / CO2}} 7. CO \texttt{\detokenize{[C]=[O]}} A major carbon-containing intermediate/product. It reaches \textasciitilde{}22 molecules and is only partially oxidized further. 8. OCO / carboxylate-like intermediate \texttt{\detokenize{[O][C]=[O]}} Appears in the CO → CO₂ oxidation pathway: \texttt{\detokenize{CO / OCO → CO2}} 9. CO₂ \texttt{\detokenize{[O]=[C]=[O]}} Final oxidation product, reaching 16 molecules by the final sampled frame. Overall, the important pathway looks like: \texttt{\detokenize{CH4 → CH3 → CH2O / CH3O-like species → CO → CO2}} coupled with oxygen radical/peroxide chemistry: \texttt{\detokenize{O2 ⇌ O–O → HO2 / H2O2 → H2O}} The most chemically important intermediates are therefore CH₃, CH₂O, HO₂, O–O/peroxide species, CO, and OCO-like species. & 2,078 & 739 & 90,624 \\
122 & \textbf{User message} & What are key intermediates in this reaction? &  &  &  \\
123 & Agent tool call & read; path: \texttt{\detokenize{methane/ch4_o2_combustion_workflow/md/ch4_o2_3000K.lammpstrj.species}}\newline read; path: \texttt{\detokenize{methane/ch4_o2_combustion_workflow/analysis/reacnet/reaction_report.md}} & 1,431 & 129 & 96,768 \\
124 & \textbf{Agent message} & The key intermediates are mainly radical/peroxide oxygen species plus partially oxidized carbon species. Most important ones from the ReacNetGenerator species/reaction outputs: | Intermediate | ReacNet label | Role | |---|---|---| | CH₃ / methyl | [H][C]([H])[H] | First methane-activation product from CH₄ C–H breaking. | | CH₂O / formaldehyde-like | [H][C]([H])=[O] | Main oxygenated C1 intermediate before CO formation. | | CH₃O / methoxy-like | [H][C]([H])[O] | Transient oxygenated hydrocarbon intermediate. | | CH₃OH / methanol-like | [H][O][C]([H])([H])[H] | Minor transient oxygenated hydrocarbon. | | HO₂ / hydroperoxyl | [H][O][O] | Important oxygen-radical intermediate; appears early and converts to H₂O/O₂/O–O species. | | H₂O₂ / peroxide-like | [H][O][O][H] | Low-count peroxide intermediate. | | O–O / activated peroxide-like oxygen | [O][O] | Very active oxygen intermediate; strong O2 ⇌ O–O cycling. | | CO | [C]=[O] | Major partially oxidized carbon product/intermediate; reaches \textasciitilde{}22 molecules. | | OCO / carboxyl-like | [O][C]=[O] | Intermediate in CO → CO₂ oxidation. | | ketene/oxygenated C₂-like fragments | [H][C]([H])=[C]=[O], [H][C]=[C]=[O] | Minor carbon-coupling/oxygenated fragments. | The clearest intermediate pathway is: \texttt{\detokenize{CH4 → CH3 → CH2O / CH3O-like species → CO → OCO → CO2}} Coupled oxygen chemistry is: \texttt{\detokenize{O2 ⇌ O–O/peroxide-like species → HO2 / H2O2 → H2O}} So, if we name only the key intermediates: CH₃, CH₂O, HO₂, O–O/peroxide species, CO, and OCO-like species. & 3,279 & 531 & 97,792 \\
\end{longtable}
\endgroup
\end{landscape}